\documentclass[eqsecnum,11pt,nofootinbib,superscriptaddress]{revtex4-1}
\pdfoutput=1

\usepackage[utf8]{inputenc}
\usepackage[T1]{fontenc}
\usepackage{microtype}
\usepackage{amsmath,amssymb,amsfonts}
\usepackage{tensor}
\usepackage{natbib}
\usepackage{graphicx}
\usepackage{dcolumn}
\usepackage{bm}
\usepackage{comment}
\usepackage{braket}
\usepackage{xcolor}
\usepackage{booktabs}
\usepackage{hyperref}
\hypersetup{
    colorlinks,
    linkcolor=black,
    citecolor=black,
    urlcolor=black,
}
\newcommand{\virgolette}[1]{``#1''}

\newcommand\appref[1]{Appendix~\ref{#1}}
\newcommand\secref[1]{Section~\ref{#1}}

\newcommand\Kappa{\mathrm{K}}

\newcommand\weakeq{\approx}

\newcommand\areaVariable{\Lambda}
\newcommand\curvatureVariable{\Kappa}
\newcommand\kaCoeff{\alpha}

\DeclareMathAlphabet{\mathup}{OT1}{\familydefault}{m}{n}

\newcommand\imagi{\mathup{i}}
\newcommand\expe{\mathup{e}}

\newcommand\mathperiod{\,.}
\newcommand\mathcomma{\,,}

\usepackage{xparse}

\NewDocumentCommand{\expval}{s m}{%
  \ensuremath{%
    \IfBooleanTF{#1}{%
      \left\langle #2 \right\rangle%
    }{%
      \langle #2 \rangle%
    }%
  }%
}

\NewDocumentCommand{\op}{s m}{%
  \ensuremath{%
    \IfBooleanTF{#1}{%
      \widehat{#2}
    }{%
      \hat{#2}
    }%
  }%
}
\NewDocumentCommand{\hcop}{s m}{%
  \ensuremath{%
    \IfBooleanTF{#1}{%
      \op*{#2}%
    }{%
      \op{#2}%
    }^{\dagger}%
  }%
}

\NewDocumentCommand{\moment}{s m}{%
  \ensuremath{%
    \Delta%
    \IfBooleanTF{#1}{%
      \left(#2\right)%
    }{%
      (#2)%
    }%
  }%
}

\NewDocumentCommand{\poissonbracket}{s m m}{%
  \ensuremath{%
    \IfBooleanTF{#1}{%
      \left\{#2, #3\right\}%
    }{%
      \{#2, #3\}%
    }%
  }%
}

\NewDocumentCommand{\commutator}{s m m}{%
  \ensuremath{%
    \IfBooleanTF{#1}{%
      \left[#2, #3\right]%
    }{%
      [#2, #3]%
    }%
  }%
}

\NewDocumentCommand{\intmeasure}{o m}{%
  \ensuremath{%
    \mathup{d}%
    \IfValueT{#1}{^{#1}}%
    #2\,%
  }%
}

\NewDocumentCommand{\compconj}{s m}{%
  \ensuremath{%
    \IfBooleanTF{#1}{%
      \overline{#2}%
    }{%
      \bar{#2}%
    }%
  }%
}

\begin{document}
\count\footins = 1000
\title{Effective cosmology from one-body operators in group field theory}
\author{Steffen Gielen}
\email{s.c.gielen@sheffield.ac.uk}
\affiliation{School of Mathematics and Statistics,
  University of Sheffield,
  Hicks Building,
  Hounsfield Road,
  Sheffield S3 7RH,
  United Kingdom}
\author{Luca Marchetti}
\email{luca.marchetti@phd.unipi.it}
\affiliation{Università di Pisa, Lungarno Antonio Pacinotti 43, 56126 Pisa, Italy, EU}
\affiliation{Arnold Sommerfeld Center for Theoretical Physics,  Ludwig-Maximilians-Universit\"at München, Theresienstrasse 37, 80333 M\"unchen, Germany, EU}
\affiliation{Istituto Nazionale di Fisica Nucleare sez.~Pisa, Largo Bruno Pontecorvo 3, 56127 Pisa, Italy, EU}
\author{Daniele Oriti}
\email{daniele.oriti@physik.lmu.de}
\affiliation{Arnold Sommerfeld Center for Theoretical Physics, Ludwig-Maximilians-Universit\"at München, Theresienstrasse 37, 80333 M\"unchen, Germany, EU}
\author{Axel Polaczek}
\email{apolaczek1@sheffield.ac.uk}
\affiliation{School of Mathematics and Statistics,
  University of Sheffield,
  Hicks Building,
  Hounsfield Road,
  Sheffield S3 7RH,
  United Kingdom}

\date{\today}

\begin{abstract}
We propose a new method for obtaining an effective Friedmann--Lema\^{i}tre--Robertson--Walker (FLRW) cosmology from the quantum gravity dynamics of group field theory (GFT), based on the idea that an FLRW universe is characterised by a few macroscopic observables. Rather than relying on assuming a particular type of quantum state and computing expectation values in such a state, here we directly start from relations between macroscopic observables (defined as one-body operators) and formulate  dynamics only for those observables. We apply the effective approach to constrained quantum systems (as developed by Bojowald and collaborators) to GFT, providing a systematic expansion in powers of $\hbar$. We obtain a kinematical phase space of expectation values and moments, which does not require an \emph{a priori} choice of clock variable. We identify a gauge fixing of the system which corresponds to choosing one of the cosmological variables (with the role of extrinsic curvature) as a clock and which allows us to rewrite the effective dynamics in relational form. We show necessary and sufficient conditions for the resulting dynamics of expectation values to be compatible with those of classical FLRW cosmology and discuss the impact of quantum fluctuations. 
\end{abstract}

\maketitle
\flushbottom

\section{Introduction}
\label{sec:introduction}

Many modern approaches to quantum gravity suggest that the classical continuum notions of space and time are emergent concepts, to be obtained from an underlying fundamental description in terms of different quantum degrees of freedom. Extracting continuum physics from such theories is generally a very complicated task, because of both conceptual and technical problems. Some of the most important outstanding issues, in relation to this task, are the problem of time, the problem of Hilbert space and the problem of the classical limit \cite{Kuchar2011,Isham1992}. There are of course many other issues, such as the implementation of diffeomorphism invariance in a discrete formalism \cite{Dittrich:2014ala} or the reliance on preferred foliations, but in the following we will focus only on these three. They are present in one form or another in any formulation of quantum gravity, but they are particularly visible in a canonical formalism.

The problem of time refers to the fact that, due to the diffeomorphism invariance of general relativity, there are no external structures with respect to which one can define (time) evolution.  As a result, the Hamiltonian of general relativity (or, more in general, of any background independent theory) is forced to vanish in absence of boundaries \cite{Rovelli:2004tv, Thiemann:2007pyv, Arnowitt:1962hi}. At the quantum level, the vanishing of the Hamiltonian results in \virgolette{frozen} dynamics, where states do not evolve in \virgolette{time}. Change and evolution can however still be defined in terms of internal physical quantities. This is the relational strategy, which is based on the use of evolving constants of motion \cite{Rovelli:1990ph,Rovelli:2001bz} (see \cite{Tambornino:2011vg} for a review), i.e., Dirac observables encoding correlations between the physical variable chosen as a clock and the remaining ones. Though these observables can be formally defined in the classical theory \cite{Dittrich:2004cb,Dittrich:2005kc}, their explicit construction, especially at the quantum level, is extremely complicated. One main reason is that one has in general very little control over the physical Hilbert space on which they should be represented. The construction of a Hilbert space by endowing the space of solutions to the quantum constraint with an appropriate inner product is a poorly understood issue; this is the Hilbert space problem mentioned above. Lastly, the problem of the classical limit refers to the extraction of (semi)classical physics from full quantum gravity, which has been often addressed through an explicit construction of specific classes of states. Semiclassicality is however not a sharp notion, and it is therefore important to clarify which results are general to semiclassical states and which ones are instead only valid for the states considered. Moreover, in approaches (like GFT or spin foam models or lattice quantum gravity) in which the basic structures of the theory are discrete, the semiclassical approximation leading to general relativity is necessarily accompanied by some form of continuum approximation, implemented via coarse graining or renormalisation group techniques. The two kinds of approximations are conceptually and technically distinct \cite{Oriti:2018tym}, albeit intertwined.

Effective techniques, describing quantum systems in terms of (a finite set of) expectation values and moments of observables associated to a given state, may play a crucial role in order to practically bypass the above difficulties, while still allowing to extract physical semiclassical predictions from the theory. These techniques have been employed already in the context of simple quantum systems for instance in \cite{Bojowald:2009zzc,Bojowald:2005cw,Bojowald:2009jj,Bojowald:2010xp,Bojowald:2010qw,Hohn:2011us}. In principle, in order to apply the framework developed in \cite{Bojowald:2009jj,Bojowald:2009zzc} to quantum gravity, it should be generalised to a field theoretic context, which will in general pose formidable technical challenges. However, in quantum gravity one is often not interested in all degrees of freedom in the theory, but rather in situations where only some degrees of freedom are relevant. In particular, the relevant physical scales where quantum gravity effects might be seen are far away from the microscopic Planck scale, and will thus involve a form of collective or coarse-grained characterisation of the fundamental quantum gravity theory. One may then develop an effective framework only for those observables which are most relevant at this coarse-grained (``hydrodynamic'') level: this extends the effective framework in a more minimal way, essentially replacing the algebra of observables of the quantum particle with a similarly finitely generated algebra of collective and macroscopic observables. Expectation values and moments of these observables are then the quantities in terms of which the effective hydrodynamic description is defined. This focus on such a coarse-grained regime can also be seen as a way to implement a continuum approximation, as part of the extraction of semiclassical physics via the chosen effective framework.

In quantum gravity, a system which can be characterised by a few collective observables is the homogeneous and isotropic FLRW universe \cite{Oriti:2016acw}. The hydrodynamic description of the fundamental quantum gravity dynamics to capture homogeneous and isotropic cosmology has been employed especially in the context of group field theories (GFTs) \cite{Gielen:2016dss}. GFTs can be seen as quantum field theories whose fundamental quanta are three-dimensional discrete chunks of space, i.e., quantum tetrahedra, whose geometric properties are encoded in group variables (these variables are equivalent to the data on open spin networks in loop quantum gravity \cite{Oriti:2013aqa}). Continuum spacetime notions are expected to emerge from these theories only after an appropriate coarse-graining is performed. Their Fock space structure is perfectly suited for the construction of non-perturbative collective states, such as coherent states, which have an interpretation in terms of continuous and homogeneous three-geometries \cite{Gielen:2013naa,Gielen:2014ila}. Using these states and mean field methods one can obtain (in the weak interaction limit) an effectively hydrodynamic description characterised by Friedmann dynamics with a quantum bounce for the averaged volume of the universe \cite{Oriti:2016qtz,Marchetti:2020umh,Gielen:2020fgi,Wilson-Ewing:2018mrp}. These results have sparked further works addressing for instance the role of quantum fluctuations, especially for late times and around the bounce \cite{Marchetti:2020qsq,Gielen:2019kae} and the definition of relational evolution in such emergent quantum gravity theories \cite{Marchetti:2020umh,Gielen:2021dlk}.

In this paper, we aim to provide a new perspective on the derivation of cosmological dynamics from GFT by extending the effective techniques developed in \cite{Bojowald:2009zzc,Bojowald:2005cw,Bojowald:2009jj} to the GFT setting. As in \cite{Oriti:2016qtz} and many subsequent papers, we consider a GFT model minimally coupled with a massless scalar field, and assume that the self-interactions of the GFT field are negligible in the regime of interest. As mentioned above, the effective techniques were developed for a quantum algebra generated by a finite set of operators. We propose 
to use a suitable finite set of one-body GFT operators\footnote{By ``one-body operator'' we mean an operator which can be written as an integral over a single combination of creation and annihilation operators.}.
 In particular, the quantum constraint of \cite{Bojowald:2009zzc} is given by the integrated (free) equation of motion in our set-up. Truncating the system in this manner is motivated by the interest in only some collective (hydrodynamic) variables, and is the most crucial approximation that we perform; we see it as a requirement to be imposed on the kind of states employed in the effective description. One important consequence of this assumption is that the number operator of GFT quanta\footnote{Note that this is the fundamental number operator and not the effective relational number operator, a function of a relational clock value, typically used in GFT cosmology \cite{Oriti:2016qtz}.} trivially commutes with the constraint, so that its expectation value and all its moments will be constant. The quantum algebra thus obtained allows to construct a quantum phase space coordinatised by expectation values and moments of the relevant operators, according to standard techniques \cite{Bojowald:2009zzc,Bojowald:2009jj,Ashtekar:1997ud}. By imposing that the states of interest are also semiclassical, one can truncate the hierarchy of moments to a desired order in $\hbar$. The quantum constraint then produces a finite set of effective constraints, generating gauge flows on the variables of the truncated quantum phase space. Similarly to \cite{Bojowald:2009jj,Bojowald:2010xp,Bojowald:2010qw}, all but one of these gauge flows are fixed so that one of the geometric operators in the algebra (here chosen to classically match the extrinsic curvature) is effectively classical, and can be used as a proper clock. Through this procedure, one ends up with a relational description of a reduced system of variables which effectively represents the physical Hilbert space.

We show how to recover the Friedmann dynamics of a flat FLRW universe at the level of expectation values. Moreover, assuming that quantum fluctuations are small at small curvature, they remain under control also large curvature, a feature shared with loop quantum cosmology (LQC) models \cite{Corichi:2011rt,Rovelli:2013zaa}. We compare these results on fluctuations with previous works \cite{Marchetti:2020qsq,Gielen:2019kae} which stressed the importance of a changing particle number in order to suppress or enhance quantum fluctuations. This is in contrast with our approach here in which the particle number is constant (keeping in mind that the notions of dynamics and time dependence differ in these different papers -- see next paragraph). Our results also provide additional insights into the role of relational dynamics in emergent quantum gravity theories, emphasising the importance of coarse-graining (in our case given by explicit truncation) before defining a notion of relationality, as suggested in \cite{Marchetti:2020umh}. 

Our method for implementing relational dynamics in GFT differs from what was done in previous papers such as  \cite{Oriti:2016qtz,Wilson-Ewing:2018mrp,Marchetti:2020umh}. More precisely, while the approach proposed in this paper shares with \cite{Marchetti:2020umh} the fact that relational evolution is defined only in an effective sense, in contrast to \cite{Marchetti:2020umh} it does not require a specific form of states. In \cite{Oriti:2016qtz}, instead, a relational notion of dynamics was implemented by defining \virgolette{relational operators} (which however need smearing in order to be well-defined \cite{Assanioussi:2020hwf,Marchetti:2020umh}), while \cite{Wilson-Ewing:2018mrp} employed a deparametrised framework from the start (the two last approaches are connected in \cite{Gielen:2021dlk}). In our approach, we do not attempt to define microscopic relational observables, nor we try to construct a reduced physical Hilbert space; these difficulties are in fact bypassed by the effective approach and the notion of relationality is defined in the usual sense of constrained dynamics on a phase space. A complete understanding of how these approaches to relational dynamics are related to each other is still missing, and certainly deserves more study.

Finally, let us stress again that in order to obtain our results one does not need to specify a state -- only general semiclassicality and \virgolette{hydrodynamic} conditions are assumed -- thus suggesting that the connection between GFTs and cosmological solutions may go beyond the choice of coherent states (as in the simpler setting of \cite{Gielen:2019kae}).

\section{Review of the effective approach}
\label{sec:effective_constraints_review}
In this section we review the effective approach to constrained systems \cite{Bojowald:2005cw, Bojowald:2009jj} and to relational dynamics \cite{Bojowald:2010xp} (see also \cite{Bojowald:2010qw,Hohn:2011us}). While there are many fascinating aspects of this approach that deserve to be discussed at length, we touch only the points necessary for the understanding of the procedure developed in this paper, referring to \cite{Bojowald:2005cw,Bojowald:2009jj,Bojowald:2010xp,Bojowald:2010qw,Hohn:2011us} for more details.
\subsection{Effective constraints}
The effective approach to quantum constrained systems is formulated in the quantum phase space, which can be constructed from first principles as follows \cite{Hohn:2012nia,Tsobanjan:2009mq}. 

The starting point is a set of elements $\hat{A}_i$ with $i=1,\dots,n$ generating an associative unital algebra $\mathcal{A}$ (which contains all finite polynomials in the $\hat{A}_{i}$) and satisfying the commutation relations
\begin{equation}\label{eq:generalfundamentalcommutations}
[\hat{A}_i,\hat{A}_j]=\imagi\hbar f\indices{_i_j^k} \hat{A}_k
\end{equation}
for some real constants $f\indices{_i_j^k}$. One usually assumes that this is an involutive algebra, i.e., there is a $*$ operation corresponding to the adjoint (where the generators are often thought of as self-adjoint).

Monomials of the form $\hat{A}_1^{m_1}\hat{A}_{2}^{m_2}\dots \hat{A}_n^{m_n}$ with $m_{i}\ge 0$ form a linear countable basis for $\mathcal{A}$. Now a quantum state  $\psi$ can be seen as a linear functional on $\mathcal{A}$, i.e., a map $\psi:\mathcal{A}\to\mathbb{C}$, with the conventional notation $\braket{\hat{A}_i}_\psi= \psi(\hat{A}_i)$ (and with $\braket{\hat{1}}_\psi=1$). From an algebraic viewpoint this is the definition of the notion of state. One connection between this algebraic framework and the canonical description in terms of operators acting on Hilbert spaces is provided by the Gelfand, Naimark and Segal (GNS) theorem \cite{gns1,gns2} according to which, given any algebraic state $\alpha$, it is always possible to construct an Hilbert space $\mathcal{H}_\alpha$ on which the algebra of observables act as an algebra of linear bounded operators $\pi_\alpha(\hat{A}_i)$, such that $\pi_\alpha$ is a $*$-representation of $\mathcal{A}$. This Hilbert space $\mathcal{H}_\alpha$ contains a cyclic vector $|\alpha\rangle\in \mathcal{H}_\alpha$ such that $\alpha(\hat{A}_i)=\langle\alpha| \pi_\alpha(\hat{A}_i)| \alpha\rangle$.
More generally, one may start from a Hilbert space $\mathcal{H}$ on which the elements in $\mathcal{A}$ act as operators; then one can again think of each $\psi\in\mathcal{H}$ as a linear functional by defining $\psi(\hat{A}_i)=\langle \psi|\hat{A}_i|\psi\rangle$. 
In either viewpoint, the statement relevant in the effective approach is that one is interested in  elements of $\bar{\mathcal{A}}$,  the dual vector space to $\mathcal{A}$. A subtle point here is that the existence of a Hilbert space representation would imply that $\braket{\hat{A}_i^*\hat{A}_i}_\psi\ge 0$ but it is mentioned in \cite{Hohn:2012nia,Tsobanjan:2009mq} that this condition need not hold in the more general algebraic framework.

Because of linearity, any $\psi\in\bar{\mathcal{A}}$ is determined by the values it assigns to a linear basis of $\mathcal{A}$. So, one could coordinatise $\bar{\mathcal{A}}$ through the \emph{expectation values} $\braket{\hat{A}_i}$ with $i=1,\dots,n$ and \emph{moments}
\begin{equation}
  \Delta(A_1^{m_1}\dots A_n^{m_n})
  =
  \left\langle
    \left(\hat{A}_1-\braket{\hat{A}_1}\right)^{m_1}
    \dots
    \left(\hat{A}_n-\braket{\hat{A}_n}\right)^{m_n}
  \right\rangle_{ \text{Weyl}}\,,
\end{equation}
where the subscript indicates a total symmetrisation of the product of operators inside the expectation value above, and where $\sum_{i=1}^nm_i\ge 2$. This choice of coordinatisation proves especially useful in effective approaches, where the states chosen are such that there is a clear hierarchy of moments, as we will discuss below.

The Poisson structure turning the state space  $\bar{\mathcal{A}}$ coordinatised by expectation values and moments into the \emph{quantum phase space} is directly inherited from the commutation relations of the elements in $\mathcal{A}$: one defines\footnote{This form of Poisson brackets can also be motivated by the geometrical formulation of quantum mechanics \cite{Ashtekar:1997ud}.}
\begin{equation}\label{eq:poissonbrackets}
\left\{\braket{\hat{A}},\braket{\hat{B}}\right\}=(\imagi\hbar)^{-1}\left\langle [\hat{A},\hat{B}]\right\rangle
\end{equation}
for any $\hat{A}$, $\hat{B}\in \mathcal{A}$ (it is enough to demand this for monomials in the generators $\hat{A}_i$). For constrained systems, the quantum phase space corresponds to a kinematical Hilbert space, since no constraint has been imposed yet. In this sense, variables in $\mathcal{A}$ are kinematical quantum variables. 

In the algebraic framework, we can now define a quantum constraint $\hat{C}$ as an algebra element satisfying\footnote{This can be seen as the algebraic counterpart of the 
Dirac condition $\hat{C}|\psi\rangle=0$ for any state $|\psi\rangle$ in a Hilbert space, the connection between algebraic and Dirac formulation being provided by the GNS construction discussed above.} $\langle\hat{A}\hat{C}\rangle=0$ for any $\hat{A}\in\mathcal{A}$ \cite{Bojowald:2009jj}. In other words, the constraint can be characterised by requiring that its expectation value and all of its fluctuations are zero:
\begin{subequations}
  \label{eq:effective_constraints}
\begin{align}
  C(\braket{\hat{A}_i},\Delta(\dots))
  &=
  \braket{\hat{C}}
  =0
  \mathcomma
  \\
  \label{Cpol-definition}
  C_{\text{pol}}(\braket{\hat{A}_i},\Delta(\dots))
  &=
  \left\langle
    \left(\widehat{\text{pol}}-\braket{\widehat{\text{pol}}}\right)\hat{C}
  \right\rangle
  =
  0
  \mathcomma
\end{align}
\end{subequations}
with $\widehat{\text{pol}}$ being a polynomial in the algebra elements. The notation here emphasises that the constraint functions should be expressed in terms of expectation values and moments\footnote{For non-polynomial constraints, this can be achieved via a Taylor expansion.}. Notice that, contrary to what is done for moments, the quantum constraints are not defined by a symmetric ordering. In turn, this implies that the quantum constraint functions can take complex values. This was however shown not to be an issue for deparametrisable systems, basically because the moments appearing in the constraints are \emph{kinematical} moments and do not need to be restricted to real values. On the other hand, physical expectation values and moments must be real\footnote{Notice that in the usual Hilbert space approach, $\hat{C}|\psi\rangle=0$ likewise only implies that $\hat{A}\hat{C}|\psi\rangle=0$ for any $\hat{A}$ but not necessarily $\left(\hat{A}\hat{C}+\hat{C}\hat{A}\right)|\psi\rangle=0$, which could lead to an analogous discussion.} \cite{Bojowald:2009jj,Bojowald:2010qw}. 

The reason for this ordering choice is that a symmetric ordering would not in general lead to a first class system of constraints \cite{Bojowald:2009jj}. On the other hand, with the definition above it is easy to show that for any $C_{\text{pol}}$ and $C_{\text{pol}'}$ defined as in (\ref{Cpol-definition}) we have
\begin{equation*}
\left\{C_\text{pol},C_{\text{pol}'}\right\}=(\imagi\hbar)^{-1}\left\langle \left[\widehat{\text{pol}}\,\hat{C},\widehat{\text{pol}}'\,\hat{C}\right]\right\rangle\approx 0
\end{equation*}
where $\approx$ is Dirac's ``weak equality'' (equality up to functions that are multiplied by constraints), and so the system closes. As a consequence, these quantum constraint functions induce \emph{quantum gauge transformations} on their solution space through their Hamiltonian flows. The presence of gauge flows even after solving the quantum constraint functions in the quantum phase space approach can be understood again as an effect of working with kinematical variables, which are indeed affected by gauge even at the classical level. Moments and expectation values of Dirac observables, instead, are gauge invariant also from this effective quantum phase space perspective. Indeed, if $\hat{O}$ is a Dirac observable, the flow of its expectation value,
\begin{equation}
  \label{eq:gauge_transformation_dirac_observable}
  \begin{aligned}
    \braket{\hat{O}}'(\tau)&=\left\{\braket{\hat{O}},\left\langle\left(\widehat{\text{pol}}-\braket{\widehat{\text{pol}}}\right)\hat{C}\right\rangle\right\}\\
    &=(\imagi\hbar)^{-1}\left[\left\langle\left(\widehat{\text{pol}}-\braket{\widehat{\text{pol}}}\right)[\hat{O},\hat{C}]\right\rangle+\left\langle[\hat{O},\widehat{\text{pol}}]\left(\hat{C}-\braket{\hat{C}}\right)\right\rangle\right]
  \end{aligned}
\end{equation}
where $\tau$ is parametrising the flow, vanishes weakly. The same of course is true for moments of Dirac observables.
In this approach, the difficulty in solving the quantum constraint and explicitly constructing a physical Hilbert space  has been traded for an infinite set of variables and constraint functions. These are in general intractable, unless one can identify some approximation that allows to reduce them to a finite set, neglecting subdominant contributions, and thus characterising the approach as an effective one. An example of such approximations is the semiclassical one, where it is assumed that moments of higher order are also higher order in $\hbar$. More precisely, for the semiclassical expansion one assumes that $\Delta(A_1^{m_1}\dots A_n^{m_n})\sim \hbar^{(m_1+\ldots+m_n)/2}$. At a given order in $\hbar$, only a finite number of constraints and moments are relevant. By construction this hierarchy is also preserved by the Poisson structure. Indeed, one can show from  \eqref{eq:poissonbrackets} that the Poisson bracket of a moment of order $ \hbar^{m/2}$ and a moment of order $ \hbar^{n/2}$ is of order $\hbar^{(m+n-2)/2}$.

In the following, we will assume that a truncation at order $\hbar$ captures the dynamics of suitable semiclassical states to a good approximation. This means that we will neglect all terms explicitly proportional to $\hbar^{3/2}$ or higher and all third or higher order moments. For an algebra generated by $n$ elements $\hat{A}_i$, we then have $n+1$ constraints $C=\braket{\hat{C}}=0$ and $C_{A_i}= \braket{(\hat{A}_i-\braket{\op{A}_i})\hat{C}}=0$ and $n+n(n+1)/2=n(n+3)/2$ expectation values and moments. 

The extension of the Poisson bracket \eqref{eq:poissonbrackets} for an algebra with commutation relations \eqref{eq:generalfundamentalcommutations} is, at the given truncation to order $\hbar$, given by the relations
\begin{subequations}
  \label{eq:poisson_brackets_truncated_quantum_phasespace}
  \begin{align}
  \poissonbracket{
    \expval{\op{A}_i}
  }{
    \moment{A_j A_k}
  }
  &
  =
  f\indices{_{ij}^l} \moment{A_l A_k}
  +
  (j \leftrightarrow k)
  \mathcomma
  \\
  \poissonbracket{
    \moment{A_i A_j}
  }{
    \moment{A_k A_l}
  }
  &
  =
  \left(
    f\indices{_i_k^m}
    \expval{\op{A}_m}
    \moment{A_j A_l}
    +
    (i \leftrightarrow j)
  \right)
  +
  (k \leftrightarrow l)
  \mathperiod
\end{align}
\end{subequations}

\subsection{Effective relational dynamics}
Within the effective framework described above, the way physical variables are selected is through an effective local deparametrisation of the system \cite{Bojowald:2009jj,Bojowald:2010qw}. This is obtained via a partial gauge fixing of the aforementioned gauge freedom (\emph{Zeitgeist}), leading to a projection into a classical \virgolette{time}, and consequently to the definition of transient relational observables called \emph{fashionables}. In \cite{Bojowald:2009jj,Bojowald:2010qw} most of the features of this effective approach to relational dynamics are explained through different (classical deparametrisable and non-deparametrisable) examples. Here we will try to provide a general summary of such features, most of which will also be true for the system we aim to study in the rest of this paper.
An important feature of the effective approach explained above is that the quantum phase space generically has a degenerate Poisson structure, i.e., is not symplectic. Indeed, moments of any arbitrary order are characterised, in general, by a degenerate Poisson tensor\footnote{As an example, consider a canonical pair $\hat{q}$, $\hat{p}$. At order $\hbar$, we have $3$ moments, so the space is odd-dimensional and the Poisson structure of second order moments must be degenerate.}.
A crucial consequence of this degeneracy of the Poisson structure is that even though the constraints $C_\text{pol}$, $C$ are independent, the gauge flows generated by them are generally not. The Poisson geometry of constrained systems is discussed in \cite{Bojowald:2001ae}.

In the examples studied in \cite{Bojowald:2010qw}, at order $\hbar$ there are $n+1$ constraints, but only $n$ gauge flows generated by them are independent. This means that of the $n(n+3)/2$ variables relevant at that order, $2n$ variables can be eliminated using the constraints ($n+1$ conditions) and $n-1$ by fixing all the gauge flows but one. This leaves $n(n+3)/2-2n=n(n-1)/2$ remaining variables. The importance of this counting can be understood when one is interested, as we have mentioned above, in an effective deparametrisation of the system. Classically, such a deparametrisation would result in the complete elimination of a (possibly canonical) pair of variables. At the effective level, eliminating all  variables related to a pair of operators having closed non-trivial commutation relations among themselves leads to a total of $(n-2)(n+1)/2=n(n-1)/2-1$ variables. So, by appropriately fixing $n-1$ gauge flows and using the $n+1$ constraints (see below), one could obtain a set of $(n-2)(n+1)/2$ physical variables and one classical \virgolette{clock variable}, all of them subject to the remaining, unfixed gauge flow. In turn, this allows constructing local relational observables (fashionables) describing the evolution of the physical variables with respect to the clock.

Let us discuss in more detail how this procedure can be realised concretely. The theory is entirely clock neutral, since all physical variables are treated on equal footing. The partial gauge fixing mentioned above, or Zeitgeist, has exactly the purpose of choosing a specific clock, projecting the quantum clock we are interested in onto a classical parameter. Of course, such a clock, say $\hat{T}$, needs to have non-trivial commutation relations with $\hat{C}$, i.e., the clock may not be a Dirac observable since otherwise it is constant (cf.\ \eqref{eq:gauge_transformation_dirac_observable}).

We will assume this, and we will again further restrict to an order $\hbar$ truncation. If the clock also forms a closed subalgebra with another operator, say $\hat{P}$ (which is the situation studied in \cite{Bojowald:2009jj,Bojowald:2010qw} and considered below), the $n-1$ conditions
\begin{equation}\label{eq:gaugefixing}
\phi_{\alpha}= \Delta(T A_\alpha)=0\,,\qquad\alpha\in \mathcal{A} \setminus \{ \hat{P} \}\,,
\end{equation}
fix $n-1$ gauge flows at order $\hbar$, and make the clock classical on the space of physical variables. Then, if the expectation value and the $n$ moments of the variable $\hat{P}$ can be eliminated using the $n+1$ effective constraints\footnote{This for instance happens if the quantum constraint is linear in the variable $\hat{P}$, which will be a case of interest in the following.}, one indeed arrives at a physical description involving the $(n-2)(n+1)/2$ expectation values and order $\hbar$ moments of the remaining $n-2$ variables, and $\braket{\hat{T}}$.

Once the gauge fixing conditions \eqref{eq:gaugefixing} are imposed, one is left with a mixture of first and second class constraints. However, since
\begin{align*}
\left\{\phi_\alpha,\phi_\beta\right\}&=\frac{1}{\imagi\hbar}\left\langle\left[(\hat{T}-\braket{\hat{T}})(\hat{A}_\alpha-\braket{\hat{A}_\alpha}),(\hat{T}-\braket{\hat{T}})(\hat{A}_\beta-\braket{\hat{A}_\beta})\right]\right\rangle\nonumber\\
&=\frac{1}{\imagi\hbar}\left\langle(\hat{T}-\braket{\hat{T}})^2\left[\hat{A}_\alpha-\braket{\hat{A}_\alpha},\hat{A}_\beta-\braket{\hat{A}_\beta}\right]\right\rangle=f\indices{_\alpha_\beta^\gamma}\left\langle(\hat{T}-\braket{\hat{T}})^2\hat{A}_\gamma\right\rangle\nonumber\\
&\simeq f\indices{_\alpha_\beta^\gamma} \braket{\hat{A}_{\gamma}}\Delta(T^2)\approx 0\,,
\end{align*}
where in the third line we have neglected higher order terms, we notice that the matrix of Poisson brackets $\Delta_{ij}=\{\chi_i,\chi_j\}$, where $\chi_i$ include first order constraints and the gauge fixing conditions \eqref{eq:gaugefixing}, is actually off-block diagonal, with non-zero values only involving physical constraints and gauge fixing conditions. This, together with the fact that the $n(n-1)/2$ remaining variables have weakly vanishing Poisson brackets with the gauge fixing constraints (which can be proven in an analogous way), shows that the Dirac brackets
\begin{equation}
\{f,g\}_{\text{Dirac}}= \{f,g\}-\{f,\chi_i\}\Delta^{ij}\{\chi_j, g\}
\,,\quad
\Delta^{ij} = (\Delta^{-1})_{ij}
\end{equation}
for the $n(n-1)/2$ remaining variables are identical to the Poisson brackets introduced above. Notice that the remaining moments must, at this point, be real, contrary to what may happen for moments of $\hat{P}$ (see for instance \cite{Bojowald:2009jj,Bojowald:2010qw}) or for the expectation value of $\hat{T}$ itself (see \cite{Bojowald:2010qw}).

Once a Zeitgeist has been chosen, for instance $T = \braket{\hat{T}}$ as specified above, one looks for the remaining gauge flow that preserves the  gauge fixing conditions and, through that, one constructs relational quantities of the form $\braket{\hat{A}_{\alpha}} \circ T^{-1}$ and $\Delta(A_\alpha A_\beta) \circ T^{-1}$. These observables are transient in nature, since they can only be meaningful as long as the clock does not reach any turning point and maintains a monotonic evolution. When this is the case, one is forced to move from one Zeitgeist to another one; correspondingly, these observables may \virgolette{fall out of fashion}. As such, these \virgolette{fashionables} can only be state dependent. We refer again to \cite{Bojowald:2010xp,Bojowald:2010qw} for more details on this topic, and \cite{Hohn:2011us} for an explicit example of a model that does not have a globally defined clock. In the case we will be interested in, we will choose a clock variable with globally monotonic evolution, so that the Zeitgeist is valid globally.

\subsection{Extending the effective approach to field theory}

The effective approach to constrained systems is usually applied to quantum-mechanical systems of a finite number of degrees of freedom, often with a single constraint $\hat{C}$. The generators $\hat{A}_i$ of $\mathcal{A}$ then usually represent the quantum analogue of phase space coordinates for a classical system one proposes to quantise (e.g., positions and momenta of a system of particles). We saw above that the quantum phase space $\bar{\mathcal{A}}$ can be identified with an (algebraically constructed) set of quantum states on the algebra $\mathcal{A}$, or with a Hilbert space that the elements of $\mathcal{A}$ could be seen as acting on.

In this paper, we want to extend the application of this approach to a field theory setting. A quantum field theory can be seen as the quantisation of a system with already classically infinitely many degrees of freedom, and would therefore be associated with a very large observable algebra $\mathcal{A}_{\text{QFT}}$ (as is the starting point, e.g., in algebraic quantum field theory). We will however restrict to a much smaller algebra representing only a few coarse-grained observables we are interested in. That is, our algebra $\mathcal{A}$ will only be a very small subalgebra of the full  $\mathcal{A}_{\text{QFT}}$. It then follows that the quantum phase space $\bar{\mathcal{A}}$ of interest to us is much smaller than the field theory quantum phase space $\bar{\mathcal{A}}_{\text{QFT}}$. The smaller  $\bar{\mathcal{A}}$ can then perhaps still be identified with some Hilbert space, but this will not be the Hilbert space of the field theory $\mathcal{H}_{\text{QFT}}$, which is much larger. Indeed there are in general different elements of $\mathcal{H}_{\text{QFT}}$ that would be identified with the same element of $\bar{\mathcal{A}}$ in the effective approach; these are states that agree on expectation values and moments of macroscopic observables even though their microscopic definition may be different\footnote{This is consistent with a coarse-graining interpretation of the reduction to an observable sub-algebra.}.

Since the effective approach as reviewed here starts from the algebra $\mathcal{A}$ and a set of constraints only involving elements in $\mathcal{A}$, the question of how the quantum phase space $\bar{\mathcal{A}}$ might sit inside a much larger Hilbert space is not of direct relevance for its application. In this sense, we will not need to give an explicit construction of $\bar{\mathcal{A}}$ as a subspace of  $\mathcal{H}_{\text{QFT}}$ which in our case will be the GFT Fock space. Our interpretation is that $\bar{\mathcal{A}}$ is a quantum phase space resulting from a coarse graining of the field theory, which leads to an effective description characterised by a few operators (among the infinitely many that can be constructed out of the field operators) whose expectation values and moments coordinatise an effective quantum phase space.

\section{Truncating the dynamics to one-body operators}
\label{sec:general-setup}

In this paper we apply the effective approach outlined in the previous section to the dynamics of group field theory, in order to obtain a systematic semiclassical expansion that does not require specifying either a particular state or a particular relational clock. Instead, we will only assume that the state is sufficiently semiclassical for an $\hbar$ expansion to make sense, and we will construct a clock using the strategy of deparametrisation and gauge fixing by a Zeitgeist explained above.

As we already outlined at the end of the previous section, out of the infinitely many independent observables that can be constructed out of the GFT field operators we will only study a few ``global'' one-body operators, and truncate the dynamics to a system of equations that only makes reference to those operators. Clearly the choice of operators in our algebra is to some extent determined by the dynamics of the theory. We implicitly assume that those few global quantities are sufficient to capture the most relevant dynamical information of the field theory. Such an assumption is conceptually similar to assuming a homogeneous cosmology in general relativity, which amounts to the assumption that a few global quantities (the volume of space, a homogeneous energy density, etc.) capture the dynamics of the Universe. Here we see this truncation as representing a coarse-grained description of the  field theory in which one is only interested in some macroscopic variables, which is justified for states for which the coarse-grained description is ``reasonably close'' to the full microscopic dynamics of the field theory. This viewpoint is then the quantum analogue of the classical perspective that a homogeneous, isotropic universe in classical cosmology is not a very symmetric solution to the Einstein equations but should rather be seen as obtained from averaging over inhomogeneities on small scales \cite{Buchert:2001sa,Buchert:2015iva}. In the setting of quantum field theory, understanding the conditions for this truncation to be valid is a more challenging problem.
\subsection{Effective GFT quantum constraint}
We now introduce the details of our effective approach to GFT dynamics. First, we introduce the type of field theories we are studying.  We consider GFT models with domain $G^4\times\mathbb{R}$, where $G$ is a Lie group representing gravitational degrees of freedom and $\mathbb{R}$ represents a scalar matter field. Writing the field as $\varphi(g_I,\chi)$ the dynamics are then schematically defined by an action
\begin{equation}
S[\varphi,\bar\varphi] = \int \intmeasure[4]{g}\intmeasure{\chi}\bar\varphi(g_I,\chi)\mathcal{D}\varphi(g_I,\chi) + I[\varphi,\bar\varphi]
\end{equation}
where $\mathcal{D}$ is a differential operator on $G^4\times\mathbb{R}$ and $I[\varphi,\bar\varphi]$ is the ``interaction'' part of the action which includes terms of higher than second order in the fields. We will focus on the free part and mostly neglect the contribution from interaction terms in the following. Such an action gives rise to equations of motion
\begin{equation}
\mathcal{D}\varphi(g_I,\chi) + \frac{\delta I[\varphi,\bar\varphi]}{\delta \bar\varphi(g_I,\chi)} = 0
\end{equation}
and the complex conjugate equation for $\bar\varphi$, but rather than solving this local equation (either at the full quantum level or in a mean-field approximation) we can multiply from the left with a conjugate field and integrate to obtain
\begin{equation}
\int \intmeasure[4]{g}\intmeasure{\chi}\bar\varphi(g_I,\chi)\mathcal{D}\varphi(g_I,\chi) + {\rm higher\; order} = 0\,,
\label{integratedeom}
\end{equation}
i.e., the quadratic part of the GFT action has to vanish if we again neglect the effect of higher order contributions. (\ref{integratedeom}), with the higher order parts assumed to be negligible, is our starting point.

We can now specify the form of $\mathcal{D}$ further in order to make the form of the resulting dynamics more explicit. A minimal non-trivial $\mathcal{D}$ would include a ``mass term'' and second derivatives with respect to the group arguments, i.e.,
\begin{equation}
\mathcal{D} = m^2 + \frac{\hbar^2}{4}\sum_{I=1}^4\Delta_{g_I} + \lambda\hbar^2\, \partial_\chi^2
\label{minimal}
\end{equation}
where $\Delta_{g_I}$ is a Laplace--Beltrami operator on $G$, acting on the argument $I$ (the reason for the factor $\frac{1}{4}$ will be clear shortly). (\ref{minimal}) is a standard form for the kinetic term of a GFT model, which has been studied in previous papers on GFT cosmology such as, e.g., \cite{Gielen:2013naa,Gielen:2016uft,deCesare:2017ynn}. It can be motivated by results from GFT renormalisation which show that Laplace--Beltrami operators with respect to group arguments are generated by radiative corrections \cite{BenGeloun:2011jnm,Carrozza:2013wda}, even if one started with a pure mass term as is often done in the spin foam literature \cite{Reisenberger:2000zc}. It could also be motivated by an effective field theory expansion in derivatives, with the further assumption that the theory should be invariant with respect to arbitrary translations in $G^4\times\mathbb{R}$ so that explicit dependence on $g_I$ or $\chi$ is excluded. We will consider (\ref{minimal}) for now but discuss possible extensions below. Notice the appearance of factors of $\hbar$ associated to each derivative, which are standard in quantum mechanics but often dropped in quantum field theory by setting $\hbar=1$. In our setting we will perform a semiclassical expansion in powers of $\hbar$ so it is important to keep these powers manifest.

The discussion so far was classical. In the quantum theory, we would expect any physical state $|\psi\rangle$ to satisfy
\begin{equation}\label{eq:linearquantumequation}
\left(m^2 + \frac{\hbar^2}{4}\sum_{I=1}^4\Delta_{g_I} + \lambda\hbar^2\, \partial_\chi^2\right)\hat\varphi(g_I,\chi)|\psi\rangle = 0
\end{equation}
but we will now replace this local condition by a much weaker, integrated version
\begin{equation}
\hat{C}|\psi\rangle = \int \intmeasure[4]{g}\intmeasure{\chi}\hat\varphi^\dagger(g_I,\chi)\left(m^2 + \frac{\hbar^2}{4}\sum_{I=1}^4\Delta_{g_I} + \lambda\hbar^2\, \partial_\chi^2\right)\hat\varphi(g_I,\chi)|\psi\rangle = 0
\label{quantumeom}
\end{equation}
which we can see as a constraint equation $\op{C} \ket{\psi} = 0$ for a certain one-body operator $\op{C}$.

It is clear that (\ref{quantumeom}) is only one out of an in principle infinitely long list of similar conditions\footnote{It can be seen as the counterpart of the infinite set of Schwinger--Dyson equations.}, obtained by integrating the local equation of motion with different operators from the left\footnote{Here, due to the linearity of \eqref{eq:linearquantumequation}, the constraints obtained by this procedure only relate operators constructed out of the same number of field operators: each integrated constraint only involves $n$-body operators with the same $n$.
}. Having made this truncation, the next step is an expansion in powers of $\hbar$ in which the still fully-quantum mechanical condition (\ref{quantumeom}) is replaced by a series of statements about expectation values. This expansion is the essence of the effective approach which we are importing into the GFT setting.

\subsection{Defining the observable algebra \texorpdfstring{$\mathcal{A}$}{A}}

In the discussion of \secref{sec:effective_constraints_review} we presented the construction of a quantum phase space as starting from an observable algebra $\mathcal{A}$ generated by a finite number of elements; quantum constraints were constructed from the elements of $\mathcal{A}$. In the field theory setting of GFT, it is advantageous to proceed in the opposite direction: having defined a constraint (\ref{quantumeom}) we now introduce a number of one-body operators whose dynamics are controlled by (\ref{quantumeom}). This set should be large enough to capture at least some physically interesting and non-trivially evolving quantities, but still small enough to keep calculations under control. We will find that this balance between making non-trivial statements about effective dynamics and keeping the complexity of the approach under control will already impose quite stringent limitations on how many quantities we can and should include. The basic operators we define are then the $\hat{A}_i$ appearing in \eqref{eq:generalfundamentalcommutations}, i.e., the generators of  $\mathcal{A}$.

The form of (\ref{quantumeom}) suggests the definitions
\begin{subequations}
\begin{align}
\hat{N} & = \int \intmeasure[4]{g}\intmeasure{\chi}\hat\varphi^\dagger(g_I,\chi)\hat\varphi(g_I,\chi)\,,\label{eqn:numberoperator}
\\\hat{\Lambda}_I & = -\hbar^2\int \intmeasure[4]{g}\intmeasure{\chi}\hat\varphi^\dagger(g_I,\chi)\Delta_{g_I}\hat\varphi(g_I,\chi)\,,
\\\hat{\Pi}_2 & = -\hbar^2\int \intmeasure[4]{g}\intmeasure{\chi}\hat\varphi^\dagger(g_I,\chi)\partial_\chi^2\hat\varphi(g_I,\chi)\,.
\end{align}
\end{subequations}
$\hat{N}$ corresponds to the GFT number operator.  As a further simplification we will assume that the state $|\psi\rangle$ we are working with satisfies $\hat{\Lambda}_I|\psi\rangle =\hat{\Lambda}_J|\psi\rangle$ for any $I$ and $J$ (which one might interpret as a notion of isotropy). We will then simply write $\hat{\Lambda}$ instead of $\hat{\Lambda}_I$, given that all equations involving any of the operators $\hat{\Lambda}_I$ will be independent of the label $I$, and the sum over $I$ will cancel the factor $\frac{1}{4}$ in (\ref{constraint}). With these considerations the quantum constraint operator defined in \eqref{quantumeom} is given by
\begin{equation}
  \label{eq:quantum_constraint_onebody}
  \op{C}
  =
  m^2 \op{N}
  - \op{\Lambda}
  - \lambda \op{\Pi}_2
  \mathperiod
\end{equation}

We will later replace the operator equation $\op{C}|\psi\rangle=0$ by a series of constraints for expectation values and higher moments, based on the effective approach introduced in Section \ref{sec:effective_constraints_review}. This set of equations corresponds to a semiclassical expansion in powers of $\hbar$. At lowest order $\hbar^0$, we have the non-trivial requirement that the sum of expectation values
\begin{equation}
C=\expval{\op{C}}=m^2\langle \hat{N}\rangle - \langle \hat{\Lambda} \rangle - \lambda \langle \hat\Pi_2 \rangle
\label{constraint}
\end{equation}
should vanish. 

The one-body operators we have defined so far all commute with each other, and thus are not governed by any non-trivial dynamics. We therefore also introduce a one-body operator
\begin{equation}
\hat{X} = \int \intmeasure[4]{g}\intmeasure{\chi}\chi\;\hat\varphi^\dagger(g_I,\chi)\hat\varphi(g_I,\chi)
\end{equation}
which we will interpret as corresponding to the value of a matter scalar field (summed over all quanta in a state, since this is an extensive quantity). If we assume the commonly used canonical commutation relations
\begin{equation}
[\hat\varphi(g_I,\chi),\hat\varphi^\dagger(g'_I,\chi')]=\delta(\chi-\chi')\int \intmeasure{h} \prod_{I=1}^4 \delta(g_I h(g_I')^{-1})
\end{equation}
for the field operators,  where the slightly unusual integration over $h$ ensures that this commutator is compatible with the gauge invariance property \cite{Gielen:2013naa}
\begin{equation}
\hat\varphi(g_I,\chi)=\hat\varphi(g_Ih,\chi)\qquad \forall\;h\in \mathup{SU}(2)\,,
\end{equation}
we find
\begin{equation}
[\hat{X},\hat{\Pi}_2]=2\imagi\hbar \left(-\imagi\hbar\int \intmeasure[4]{g} \intmeasure{\chi}\hat\varphi^\dagger(g_I,\chi)\partial_\chi\hat\varphi(g_I,\chi)\right)= 2\imagi\hbar\,\hat{\Pi}_1
\end{equation}
so that $\hat{\Pi}_1$ defined by this equation also needs to be added to our algebra of one-body operators. We then also find
\begin{equation}
[\hat{X},\hat{\Pi}_1]=\imagi\hbar\,\hat{N}
\end{equation}
while all other commutators among quantities we have considered so far (including all commutators of $\hat{N}$ or $\hat{\Lambda}$) are zero. We also add a one-body operator $\hat{\Kappa}$ which has a non-trivial commutator with $\hat{\Lambda}$ but commutes with all other operators, such that the full set of operators forms a closed algebra. We will interpret the operator $\hat{\Kappa}$ as corresponding to a notion of extrinsic curvature. The precise form of the commutator is unspecified for now; later on, we will constrain it by physical arguments and find explicit realisations of such an operator in particular examples.

The six one-body operators we have defined provide the basis for the extraction of effective cosmological dynamics. The operators are subject to a single constraint (\ref{eq:quantum_constraint_onebody}).  In general, it is easy to find examples of systems of one-body operators subject to more than one constraint\footnote{For instance, consider a hypothetical GFT model with quantum field equation $(f(g_I,\chi)+1)\hat\varphi(g_I,\chi)=0$, and consider the operators $\hat{N}$, $\hat{F}_+= \int \intmeasure[4]{g}\intmeasure{\chi}\hat{\varphi}^\dagger(g_I,\chi)f(g_I,\chi)\hat{\varphi}(g_I,\chi)$ and $\hat{F}_-= \int \intmeasure[4]{g}\intmeasure{\chi}\hat{\varphi}^\dagger(g_I,\chi)f^{-1}(g_I,\chi)\hat{\varphi}(g_I,\chi)$. The equations of motion would then imply two separate constraints  $\hat{F}_++\hat{N}=0$ and $\hat{N}+\hat{F}_-=0$.}, but the specific form of the equations of motion and the algebra defined above guarantee that \eqref{eq:quantum_constraint_onebody} is the only constraint that involves operators in the algebra alone. Indeed, consider an action on \eqref{eq:linearquantumequation} from the left resulting in an integrated constraint which features an operator included in the above algebra but different from $\hat{N}$, $\hat{\Lambda}$, and $\hat{\Pi}_2$ (e.g., $\hat{\Kappa}$).
This new constraint would inevitably involve also new operators whose matrix elements depend both on the group and the scalar field variables, which are however not included in the algebra of operators of interest.  In this sense, we are considering a minimal self-consistent set of operators and dynamical relations.

Notice again that in this approach there is no dependence on any time parameter, nor a viewpoint that any of these operators define by themselves a notion of relational observable. Any notion of relational dynamics will only arise from comparing the dynamics of these operators to each other. Importantly, and in line with the philosophy of the effective approach, we make no explicit assumption about the quantum state, only that it is sufficiently semiclassical (with respect to these global observables). 
We do however see our approximation (of not solving local field equations but only demanding that our one-body operators satisfy certain ``averaged'' relations) as an implicit condition on the quantum state.

\subsection{Fixing the free commutator by analogy with classical Friedmann cosmology}
\label{sec:relation_to_classical_friedmann_cosmology}

The lowest-order truncation to the full quantum constraint would be to only require that the expectation value of the constraint vanishes. This leads to the constraint
\begin{equation}
  \label{againtheconstraint}
  C
  =
  m^2\expval{\op{N}} - \expval{\op{\areaVariable}} - \lambda \expval{\op{\Pi}_2}
  \weakeq
  0
\end{equation}
that we already defined in (\ref{constraint}). At the same level of truncation, the dynamics of our six quantum one-body operators can be reduced to the dynamics of the expectation values of these operators. The resulting dynamical system is then equivalent to a classical constrained system: the expectation values represent phase space variables subject to a single Hamiltonian constraint. The constraint generates gauge transformations, which we will assume can be interpreted as reparametrisation in time or as the time evolution flow generated by a Hamiltonian built from this constraint.

We can then demand as a minimal requirement that this system corresponds to the dynamics of a homogeneous isotropic Friedmann--Lemaître--Robertson--Walker (FLRW) universe in general relativity or a modified version of it which includes higher-curvature corrections. We will use this minimal requirement to fix the so far unspecified commutator between $\op{\areaVariable}$ and $\op{\curvatureVariable}$.

The flow generated by (\ref{againtheconstraint}) is
\begin{equation}
  \label{flow1}
  \expval{\op{N}}'(t)
  =
  \expval{\op{\Pi}_1}'(t)
  =
  \expval{\op{\Pi}_2}'(t)
  =
  \expval{\op{\areaVariable}}'(t)
  =
  0
\end{equation}
so that we immediately have four constants of motion, and
\begin{equation}
  \label{flow2}
  \expval{\op{X}}'(t)
  =
  - 2\lambda\expval{\op{\Pi}_1}(t)
  \mathcomma\quad
  \expval{\op{\curvatureVariable}}'(t)
  =
  \poissonbracket*{
    \expval{\op{\areaVariable}}
  }{
    \expval{\op{\curvatureVariable}}
  }(t)
  \mathperiod
\end{equation}
As we have stated when introducing it, the variable $\op{X}$ represents a matter scalar field whose value is summed over all quanta in a GFT state. Normally we would have to worry about the fact that this is an extensive variable and perhaps divide by the average particle number to obtain an intrinsic notion of matter scalar field. In our framework here, the variable $\expval{\op{N}}$ is always a constant of motion, since the operator $\op{N}$ commutes with all other one-body operators we consider, hence we do (for now) not have to worry about the distinction between intensive and extensive quantities.

An immediate consequence from (\ref{flow1}) and (\ref{flow2}) is then that the evolution of $\expval{\op{X}}$ is monotonic if we assume that the expectation value $\expval{\op{\Pi}_1}$ does not vanish, indeed we have
\begin{equation}
  \expval{\op{X}}(t)
  =
  \expval{\op{X}}(0) - 2\lambda\expval{\op{\Pi}_1}(0)\,t
  \mathperiod
\end{equation}
$\expval{\op{X}}$ can then be seen as a good clock, very similar to the way massless scalar fields often appear as clocks in quantum gravity and quantum cosmology.

Let us now compare this system explicitly with the dynamics of a flat FLRW universe filled with a massless scalar field in general relativity. For this system, the Hamiltonian is (see, e.g., \cite{BojowaldBook})
\begin{equation}
  \mathcal{H}_{\text{GR}}
  =
  N \left(-\frac{2\pi G}{3}\frac{p_a^2}{a}+\frac{p_\varphi^2}{2a^3}\right)
\end{equation}
where $a$ is the scale factor, $p_a$ is its canonical momentum and $p_\varphi$ is the conjugate momentum of the scalar field $\varphi$. A convenient lapse choice, $N=a^3$, transforms this to the Hamiltonian
\begin{equation}
  \label{eq:GRHamiltonian}
  \mathcal{H}_{\text{GR}}\big|_{N=V}
  =
  -6\pi G V^2 p_V^2+\frac{p_\varphi^2}{2}
\end{equation}
where we switch to a volume variable $V=a^3$ with conjugate momentum $p_V$. The equation of motion for the scalar field is now $\dot\varphi=p_\varphi$ and $p_\varphi$ is a constant of motion, a similar form to the first equation in (\ref{flow2}). This similarity would motivate further the identification of $\expval{\op{X}}$ as corresponding to a matter scalar field. How about the equations for the spacetime geometry? We have
\begin{equation}
  \dot{V}
  =
  \left\{V,\mathcal{H}_{\text{GR}}\big|_{N=V}\right\}
  =
  -12\pi G (V p_V) V
\end{equation}
so that, taking into account that $V p_V$ is clearly a constant of motion, on-shell we have $\dot{V}= cV$ for some constant of motion $c$, leading to exponential solutions. Very similarly we have
\begin{equation}
  \label{PVdynamics}
  \dot{p}_V
  =
  \poissonbracket*{
    p_V
  }{
    \mathcal{H}_{\text{GR}}\big|_{N=V}
  }
  =
  12\pi G (V p_V) p_V
\end{equation}
and again exponential solutions (with inverse sign in the exponent, again ensuring that $V p_V$ is a constant of motion). Notice that from the vanishing of \eqref{eq:GRHamiltonian} we have $V p_V = \pm p_\varphi/\sqrt{12\pi G}$ where the sign is determined by initial conditions.

The explicit solutions for the scalar field $\varphi$ and extrinsic curvature $p_V$ are then given by
\begin{subequations}
  \begin{align}
    \varphi(t)
    &
    =
    p_\varphi t
    +
    \varphi(0)
    \mathcomma
    \\
    p_V(t)
    &
    =
    p_V(0)
    \exp(
      \pm\sqrt{12\pi G}\,p_{\varphi} t
    )
    \mathcomma
  \end{align}
\end{subequations}
where $p_\varphi$ and the sign $\pm$ are constants of motion. This explicit form shows in particular that $\varphi$ and $p_V$ are globally monotonic functions, and so either of them could be used as a clock for the other degrees of freedom. If we are only interested in comparing the variables $\varphi$ and $p_V$ to each other, we can give the relation between the two as
\begin{subequations}
  \label{eq:flrw_relational_dynamics}
  \begin{align}
    (\varphi \circ p_V^{-1})(p_V)
    &
    =
    \pm\frac{1}{\sqrt{12\pi G}}
    \log\left(
      \frac{p_V}{p_V(0)}
    \right)
    +
    \varphi(0)
    \mathcomma
    \\
    \label{eq:pvexponential}
    (p_V \circ \varphi^{-1})(\varphi)
    &
    =
    p_V(0)
    \exp\left(
      \pm \sqrt{12\pi G}
      (\varphi-\varphi(0))
    \right)
    \mathperiod
  \end{align}
\end{subequations}
Notice that $p_V$ cannot change sign, so the logarithm in the first relation is always well-defined.

We now want to reproduce these classical cosmological solutions, expressed in relational language. Comparing (\ref{PVdynamics}) with the second equation in (\ref{flow2}), if we assume that the quantity $\expval{\op{\curvatureVariable}}$ is analogous to $p_V$ in the classical cosmological model, and hence represents extrinsic curvature, dynamics analogous to those of general relativity will be obtained from assuming that
\begin{equation}
  \poissonbracket*{
    \expval{\op{\areaVariable}}
  }{
    \expval{\op{\curvatureVariable}}
  }
  \propto
  \expval{\op{\curvatureVariable}}
  \mathperiod
\label{newcommutator}
\end{equation}
In the example of general relativity, this proportionality is only given on-shell (identifying the phase space variable $V p_V$ with a conserved quantity) so \emph{a priori} it would make sense to assume that also (\ref{newcommutator}) only holds on-shell. However, the basic premise of our effective approach is to assume that the commutator of two elementary variables is a linear combination of elementary variables, as in (\ref{eq:generalfundamentalcommutations}). The proportionality factor in (\ref{newcommutator}) then needs to be just a numerical constant, which we will denote by $\alpha$; we then add the commutator
\begin{equation}
  \label{areacurvcomm}
  \commutator{
    \op{\areaVariable}
  }{
    \op{\curvatureVariable}
  }
  =
  \imagi\hbar\,\kaCoeff \op{\curvatureVariable}
\end{equation}
to our algebra of basic operators. This commutator would also suggest that $\expval{\op{\areaVariable}}$ is analogous to $V p_V$ in the classical theory, consistent with the fact that $\expval{\op{\areaVariable}}$ is constant (cf.~(\ref{flow1})).

For completeness, we also discuss the most general case in which the commutator takes the form
\begin{equation}
  \label{eq:commutator_area_curvature_linear_combination}
  \commutator{
    \op{\areaVariable}
  }{
    \op{\curvatureVariable}
  }
  =
  \imagi \hbar
  \sum_{a \in \mathcal{V}}
  \alpha_a
  \op{a}
\end{equation}
where $\mathcal{V}$ is a set of labels (\virgolette{variables}),
$\mathcal{V} = \{{N}, {\Pi}_1, {\Pi}_2, {X}, {\areaVariable}, {\curvatureVariable}\}$. Interestingly, even this most general form can be solved exactly, with the solution to \eqref{flow2} in the case of $\alpha_{\curvatureVariable} \neq 0$ given by
\begin{equation}
  \label{eq:generalksolution}
  \expval{\op{\curvatureVariable}}(t)
  =
  \left(
    \frac{x}{\alpha_\curvatureVariable}
    -
    \frac{\alpha_X}{\alpha_\curvatureVariable^2}
    2 \lambda \expval{\op{\Pi}_1}(0)
  \right)
  (\expe^{\alpha_{\curvatureVariable} t} - 1)
  +
  \expval{\op{\curvatureVariable}}(0)
  \expe^{\alpha_{\curvatureVariable} t}
  + \frac{\alpha_{X}}{\alpha_{\curvatureVariable}} 2 \lambda \expval{\op{\Pi}_1}(0) t
\end{equation}
and in the case $\alpha_{\curvatureVariable} = 0$ by the expression
\begin{equation}
  \expval{\op{\curvatureVariable}}(t)
  =
  \expval{\op{\curvatureVariable}}(0)
  +
  x t
  -
  \alpha_X
  \lambda
  \expval{\op{\Pi}_1}(0)
  t^2
  \mathcomma
\end{equation}
where in both cases we defined
\begin{equation}
  x
  =
  \alpha_{N}
  \expval{\op{N}}(0)
  +
  \alpha_{\Pi_1}
  \expval{\op{\Pi}_1}(0)
  +
  \alpha_{\Pi_2}
  \expval{\op{\Pi}_2}(0)
  +
  \alpha_{X}
  \expval{\op{X}}(0)
  +
  \alpha_{\areaVariable}
  \expval{\op{\areaVariable}}(0)
  \mathperiod
\end{equation}
This result would suggest that a more general form of the commutator between $\op{\areaVariable}$ and $\op{\curvatureVariable}$ could lead to a modified effective cosmology, different from the classical case. In principle, this is  what we would expect: the effective cosmology of a quantum gravity theory such as GFT does not need to exactly reproduce classical Friedmann cosmology, and high-energy corrections such as those leading to a bounce \cite{Oriti:2016qtz,Oriti:2016ueo,Gielen:2019kae} are expected and desirable. In the closely related setting of loop quantum cosmology, the expression for the extrinsic curvature also receives such corrections compared to the classical case, and is proportional to $\arctan(\exp(\pm\sqrt{12\pi G}\phi))$ rather than the classical exponential \eqref{eq:pvexponential} \cite{Ashtekar:2011ni,Calcagni:2012vb}, and these corrections are responsible for the resolution of the cosmological singularity.

However, if we at least require the resulting effective dynamics to reduce to classical cosmology when curvature is low, so that there is a regime for $t$ in which $\expval{\op{\curvatureVariable}}(t)$ asymptotically approaches $\expe^{\alpha_{\curvatureVariable} t}\ll 1$, we do obtain quite stringent constraints from (\ref{eq:generalksolution}): we must have
\begin{equation}
  x
  =
  \alpha_X
  = 
  0
  \mathperiod
\end{equation}
The first condition $x=0$ can be satisfied by fine-tuning initial conditions or, if one requires it to hold for generic initial states, by setting 
\begin{equation}
  \alpha_{N}
  =
  \alpha_{\Pi_1}
  =
  \alpha_{\Pi_2}
  =
  \alpha_{\areaVariable}    
  =
  0
\end{equation}
reducing us to the case (\ref{areacurvcomm}), which we will therefore assume in the following. It would be interesting to study in future work whether the more general solution (\ref{eq:generalksolution}) can be given a physical interpretation perhaps in terms of a modified theory of gravity.

We conclude this section with a few remarks about the operator $\op{\curvatureVariable}$.
Firstly, the above discussion tentatively puts forward an identification of $\op{\curvatureVariable}$ with the extrinsic curvature.
However, it should also be possible to motivate the operator representing extrinsic curvature by considering the microscopic degrees of freedom.
Note that such a motivation from microscopic degrees of freedom has been employed previously in GFT where the definition of the volume operator has been imported from loop quantum gravity (LQG) \cite{Oriti:2016qtz,Oriti:2016ueo}.
In our framework this would mean that we are able to specify the integral kernel in the definition
\begin{equation}
  \op{\curvatureVariable}
  =
  \int
  \intmeasure[4]{g}
  \intmeasure[4]{g'}
  \intmeasure{\chi}
  \intmeasure{\chi'}
  \hcop{\varphi}(g_I, \chi)
  \curvatureVariable(g_I, g'_I, \chi, \chi')
  \op{\varphi}(g'_I, \chi')
  \mathperiod
\end{equation}
Since such one-body operators describe local quantities from the perspective of an LQG spin network, the identification of curvature degrees of freedom is not  straightforward (see, e.g., the discussion in \cite{Gielen:2013naa}). A starting point could be the \emph{twisted geometries} parametrisation of the LQG phase space, which identifies a discrete analogue of extrinsic curvature for each spin network link \cite{Freidel:2010aq}.

Secondly, it is not possible to find operators representing the algebra \eqref{areacurvcomm} on compact spaces.
To see that this is true, note that on a compact space the eigenbasis of the Laplacian will be countable.
Assume that we have Hermitian matrices $\curvatureVariable_{ij}$ and $\areaVariable_{ij} = \areaVariable_i \delta_{ij}$ representing the algebra on this countable eigenbasis of the Laplacian.
Then the commutator gives
\begin{equation}
  (
    \areaVariable_i
    -
    \areaVariable_j
  )
  \curvatureVariable_{ij}
  =
  \imagi\hbar\,
  \kaCoeff
  \curvatureVariable_{ij}
\end{equation}
which implies that $\curvatureVariable_{ij} = 0$. In our GFT setting, this argument would imply that a representation of 
\eqref{areacurvcomm} only exists if the Lie group $G$ chosen to define the GFT model is non-compact.

Let us stress, however, how crucial the specific form of the microscopic dynamics is also for this interpretation of $\op{\curvatureVariable}$ as an extrinsic curvature operator. Indeed, as an example consider, instead of \eqref{eq:linearquantumequation}, the constraint equation
\begin{equation}
  \left(
    m(g_I)^2
    +
    \frac{\hbar^2}{4}
    \sum_{I=1}^4
    \Delta_{g_I}
    +
    \lambda(g_I)
    \hbar^2
    \partial^2_\chi
  \right)
  \op{\varphi}(g_I,\chi)
  \ket{\psi}
  =
  0
  \mathcomma
\end{equation}
where both $m(g_I)^2$ and $\lambda(g_I)$ are non-trivial\footnote{The dependence of $m^2$ and $\lambda$ on the group variables is generally expected for GFTs constructed by path-integral identification with simplicial gravity models minimally coupled with a free massless scalar field \cite{Li:2017uao}. This dependence actually encodes the non-trivial coupling between matter and geometry degrees of freedom.} and non-singular functions of the group variables, such that their ratio is a constant, $m(g_I)^2/\lambda(g_I)= \tilde{\lambda}$. We could then consider the one-body quantum constraint
\begin{equation}
  \op{\tilde{C}}
  =
  \op{N}
  - \op{\tilde{\Lambda}}
  - \tilde{\lambda} \op{\Pi}_2
\end{equation}
with
\begin{equation}
  \op{\tilde{\Lambda}}
  =
  - \hbar^2
  \int
  \intmeasure[4]{g}
  \intmeasure{\chi}
  \hcop{\varphi}(g_I,\chi)
  m(g_I)^{-2}
  \Delta_{g_I}
  \op{\varphi}(g_I,\chi)
  \mathperiod
\end{equation}
The dynamics generated by the expectation value of the above constraint would be the same as before (putting $m^2=1$), thus implying that an operator $\op{\tilde{\Kappa}}$ such that 
\begin{equation}
\left[\hat{\tilde{\Lambda}},\hat{\tilde{\curvatureVariable}}\right]=\imagi\hbar\alpha\,\hat{\tilde{\curvatureVariable}}
\end{equation}
could be interpreted classically as extrinsic curvature. This operator, however, would be substantially different from $\hat{\Kappa}$ introduced above: since $\hat{\tilde{\Lambda}}$ would not be diagonal on the Laplacian eigenbasis, one could not exclude a representation of $\hat{\tilde{\curvatureVariable}}$ on a compact Lie group $G$ on the basis of the argument given above for $\hat{\curvatureVariable}$.

This should make clear that the precise geometric interpretation of both fundamental and effective observables is a non-trivial matter and should be taken with some caution.

\section{One-body relational cosmology}
\label{sec:one-body_relational_cosmology}
We now apply the effective approach reviewed in \secref{sec:effective_constraints_review} to the one-body group field theory model introduced in \secref{sec:general-setup}%
\footnote{%
  We performed the calculation using Wolfram Mathematica.
  A Mathematica package and a program containing the calculations of this section are available at \url{https://gitlab.com/qggftc/effective-constraints}.
}.
The quantum constraint is given in \eqref{eq:quantum_constraint_onebody},
\begin{equation}
  \label{eq:quantum_constraint_onebody_again}
  \op{C}
  =
  m^2 \op{N}
  - \op{\areaVariable}
  - \lambda \op{\Pi}_2
  \mathcomma
\end{equation}
and the quantum algebra is generated by
$\mathcal{A} = \{\op{a}\}_{a \in \mathcal{V}}$,
where $\mathcal{V}$ is the set of labels introduced in \eqref{eq:commutator_area_curvature_linear_combination}.
In the following it sometimes proves useful to make an identification $\Pi_0 = N$ allowing for a very compact notation.
This further necessitates the introduction of the symbol $\Pi_{-1}$ which however will always be multiplied by zero.

Following the arguments from classical cosmology, we now assume that the non-trivial commutation relations of the algebra are given by
\begin{subequations}
\label{algebra}
  \begin{align}
    \commutator{\op{X}}{\op{\Pi}_1}
    &
    =
    \imagi \hbar \op{N}
    \mathcomma
    \\
    \commutator{\op{X}}{\op{\Pi}_2}
    &
    =
    \imagi \hbar \, 2 \op{\Pi}_1
    \mathcomma
    \\
    \commutator{\op{\areaVariable}}{\op{\curvatureVariable}}
    &
    =
    \imagi \hbar \, \kaCoeff \op{\curvatureVariable}
    \mathperiod
  \end{align}
\end{subequations}
We will assume that an expansion in orders of moments corresponds to an expansion in orders of $\hbar$, as anticipated, and truncate the quantum phase space at second order moments.
The resulting quantum phase space is therefore $27$-dimensional, consisting of $6$ expectation values and $21$ moments.
The Poisson structure on the quantum phase space is given by the Poisson bracket induced by the commutator as in \eqref{eq:poissonbrackets}.
We collect the explicit form of the resulting Poisson structure in \appref{app:quantum_poisson_structure}.
The $7$ effective constraints, as defined in \eqref{eq:effective_constraints}, are given by
\begin{subequations}
  \label{eq:effective_constraints_one_body}
  \begin{align}
    C
    &
    =
    m^2 \expval{\op{N}}
    - \expval{\op{\areaVariable}}
    -\lambda  \expval{\op{\Pi}_2}
    \mathcomma
    \\
    C_N
    &
    =
    m^2 \moment{N^2}
    - \moment{N \areaVariable}
    - \lambda  \moment{N \Pi_2}
    \mathcomma
    \\
    C_{\Pi_1}
    &
    =
    m^2 \moment{N \Pi_1}
    - \moment{\Pi_1 \areaVariable}
    - \lambda  \moment{\Pi_1 \Pi_2}
    \mathcomma
    \\
    C_{\Pi_2}
    &
    =
    m^2 \moment{N \Pi_2}
    - \moment{\Pi_2 \areaVariable}
    - \lambda  \moment{\Pi_2^2}
    \mathcomma
    \\
    C_X
    &
    =
    m^2 \moment{N X}
    - \moment{X \areaVariable}
    - \imagi \hbar \lambda \expval{\op{\Pi}_1}
    - \lambda \moment{\Pi_2 X}
    \mathcomma
    \\
    C_{\areaVariable}
    &
    =
    m^2 \moment{N \areaVariable}
    - \moment{\areaVariable^2}
    - \lambda  \moment{\Pi_2 \areaVariable}
    \mathcomma
    \\
    C_{\curvatureVariable}
    &
    =
    m^2 \moment{N \curvatureVariable}
    + \frac{\imagi \hbar}{2} \kaCoeff \expval{\op{\curvatureVariable}}
    - \moment{\areaVariable \curvatureVariable}
    - \lambda  \moment{\Pi_2 \curvatureVariable}
    \mathperiod
  \end{align}
\end{subequations}

Although the algebra considered here does not have any canonically conjugate pair, we will nevertheless closely mimic the treatment of a system with a canonical pair as discussed in \secref{sec:effective_constraints_review}.
As a \virgolette{clock} we choose the operator $\op{\curvatureVariable}$ and as the \virgolette{conjugate} we choose $\op{\areaVariable}$.
From the effective constraints $C$ and $C_a$, $a \in \mathcal{V}$, it is then possible to eliminate the expectation value $\expval{\op{\areaVariable}}$ and the $6$ moments $\Delta(\areaVariable a)$, $a \in \mathcal{V}$.
Likewise, we choose a gauge in which all but one of the moments of the operator $\op{\curvatureVariable}$ vanish.
This can be achieved by the gauge fixing conditions
\begin{equation}
  \label{eq:gauge_fixing_one_body}
  G_a = \Delta(\curvatureVariable a) = 0
  \mathcomma
  \quad
  a \in \mathcal{V} \setminus \{\areaVariable\}
  \mathperiod
\end{equation}
To see that this does fix the gauge sufficiently, we compute the following Poisson brackets for any $a \in \mathcal{V} \setminus \{ \areaVariable \}$
(with the identification $\Pi_0 = N$),
\begin{subequations}
  \label{eq:poisson_brackets_gauge_fixing}
  \begin{align}
    \poissonbracket{G_a}{C}
    &
    \weakeq
    0
    \mathcomma
    \\
    \poissonbracket{G_a}{C_{\Pi_n}}
    &
    \weakeq
    (1 - \delta_{a \curvatureVariable})
    \kaCoeff
    \expval{\op{\curvatureVariable}}
    \left(
      \moment{\Pi_n a}
      -
      \delta_{a X}
      \frac{\imagi \hbar}{2}
      n
      \expval{\op{\Pi}_{n-1}}
    \right)
    \\
    \poissonbracket{G_a}{C_X}
    &
    \weakeq
    (1 - \delta_{a \curvatureVariable})
    \kaCoeff
    \expval{\op{\curvatureVariable}}
    \left(
      \moment{X a}
      +
      \delta_{a \Pi_n}
      \frac{\imagi\hbar}{2}
      n
      \expval{\op{\Pi}_{n-1}}
    \right)
    \mathcomma
    \\
    \poissonbracket{G_{\Pi_n}}{C_\areaVariable}
    &
    \weakeq
    \kaCoeff
    \expval{\op{\curvatureVariable}}
    \left(
      m^2 \moment{N \Pi_n}
      - \lambda \moment{\Pi_2 \Pi_n}
    \right)
    \mathcomma
    \\
    \poissonbracket{G_{X}}{C_\areaVariable}
    &
    \weakeq
    \kaCoeff
    \expval{\op{\curvatureVariable}}
    \left(
      m^2 \moment{N X}
      - \lambda \moment{\Pi_2 X}
      - 3 \imagi \hbar \lambda \expval{\op{\Pi}_1}
    \right)
    \mathcomma
    \\
    \poissonbracket{G_{\curvatureVariable}}{C_\areaVariable}
    &
    \weakeq
    2 \imagi \hbar
    \kaCoeff^2
    \expval{\op{\curvatureVariable}}^2
    \mathcomma
    \\
    \poissonbracket{G_a}{C_\curvatureVariable}
    &
    \weakeq
    0
    \mathperiod
  \end{align}
\end{subequations}
From this we see that both $C$ and $C_\curvatureVariable$ remain unfixed by our choice of gauge.
However, it turns out that only $C$ has a non-trivial flow on the gauge-fixed constraint hypersurface.
Therefore, we may study the flow generated by $C$ on the gauge-fixed constraint hypersurface and interpret this as the dynamics of the system.
After imposing the $7$ effective constraints (cf.~\eqref{eq:effective_constraints_one_body}) and the $5$ gauge fixing conditions (cf.~\eqref{eq:gauge_fixing_one_body}), the reduced quantum phase space has coordinates given by the $27 - 7 - 5 = 15$ variables
$
\tilde{\mathcal{V}}
=
\{
  \expval{\op{N}},\allowbreak
  \expval{\op{\Pi}_1},\allowbreak
  \expval{\op{\Pi}_2},\allowbreak
  \expval{\op{X}},\allowbreak
  \expval{\op{\curvatureVariable}},\allowbreak
  \moment{N ^2},\allowbreak
  \moment{N \Pi_1},\allowbreak
  \moment{\Pi_1^2},\allowbreak
  \moment{N \Pi_2},\allowbreak
  \moment{\Pi_1 \Pi_2},\allowbreak
  \moment{\Pi_2^2},\allowbreak
  \moment{N X},\allowbreak
  \moment{\Pi_1 X},\allowbreak
  \moment{\Pi_2 X},\allowbreak
  \moment{X^2}
\}
$.

We now study the flow generated by the constraint which has not been fixed by our choice of gauge.
That is, we are interested in solving the set of differential equations
\begin{equation}
  \label{eq:eom_reduced_phase_space}
  a'(t)
  =
  \poissonbracket{a}{C}(t)
  \mathcomma
  \quad
  a \in \tilde{\mathcal{V}}
  \mathperiod
\end{equation}
The equations with non-vanishing right-hand side are given by
\begin{subequations}
  \label{eq:eom_reduced_phase_space_non-trivial}
  \begin{align}
    \expval{\op{X}}'(t)
    &
    =
    -2 \lambda  \expval{\op{\Pi}_1}(t)
    \mathcomma
    \\
    \expval{\op{\curvatureVariable}}'(t)
    &
    =
    \kaCoeff \expval{\op{\curvatureVariable}}(t)
    \mathcomma
    \\
    \moment{N X}'(t)
    &
    =
    -2 \lambda  \moment{N \Pi_1}(t)
    \mathcomma
    \\
    \moment{\Pi_1 X}'(t)
    &
    =
    -2 \lambda  \moment{\Pi_1^2}(t)
    \mathcomma
    \\
    \moment{\Pi_2 X}'(t)
    &
    =
    -2 \lambda  \moment{\Pi_1 \Pi_2}(t)
    \mathcomma
    \\
    \moment{X^2}'(t)
    &
    =
    -4 \lambda  \moment{\Pi_1 X}(t)
    \mathperiod
  \end{align}
\end{subequations}
The variables which are non-constant under the gauge flow have the following dependence on the gauge flow parameter:
\begin{subequations}
  \begin{align}
    \label{eq:flow_expval_scalar_field}
    \expval{\op{X}}(t)
    &
    =
    \expval{\op{X}}(0)
    - 2 \lambda \expval{\op{\Pi}_1}(0) t
    \mathcomma
    \\
    \label{eq:flow_expval_curvature_variable}
    \expval{\op{\curvatureVariable}}(t)
    &
    =
    \expval{\op{\curvatureVariable}}(0) \expe^{\kaCoeff t}
    \mathcomma
    \\
    \moment{N X}(t)
    &
    =
    \moment{N X}(0)
    - 2 \lambda \moment{N \Pi_1}(0) t
    \mathcomma
    \\
    \moment{\Pi_1 X}(t)
    &
    =
    \moment{\Pi_1 X}(0)
    - 2 \lambda \moment{\Pi_1^2}(0) t
    \mathcomma
    \\
    \label{eq:flow_Pi2X_Momentum}
    \moment{\Pi_2 X}(t)
    &
    =
    \moment{\Pi_2 X}(0)
    - 2 \lambda \moment{\Pi_1 \Pi_2}(0) t
    \mathcomma
    \\
    \moment{X^2}(t)
    &
    =
    \moment{X^2}(0)
    - 4 \lambda \moment{\Pi_1 X}(0) t
    + 4 \lambda ^2 \moment{\Pi_1^2}(0) t^2
    \mathperiod
  \end{align}
\end{subequations}
All the remaining variables remain constant along the flow.
It is interesting to note that both \eqref{eq:flow_expval_scalar_field} and \eqref{eq:flow_expval_curvature_variable} can be inverted.
Hence both $\expval{\op{X}}$ and $\expval{\op{\curvatureVariable}}$ could be viewed as a \virgolette{clock} from a relational point of view, in the sense that their values can be used to parametrise all other dynamical variables. Of course, there is still a difference between the two since we used a gauge fixing of moments following the perspective presented in  \secref{sec:effective_constraints_review} in which $\expval{\op{\curvatureVariable}}$ is seen as the clock, so that in our gauge the variable $X$ has non-trivial quantum fluctuations. If we want to interpret \eqref{eq:flow_expval_scalar_field} and \eqref{eq:flow_expval_curvature_variable} as representing effective cosmological dynamics constructed from our set of one-body operators, which would be analogous to the classical Friedmann cosmology discussed in \secref{sec:relation_to_classical_friedmann_cosmology}, we could  view either $\expval{\op{X}}$ as a function of $\expval{\op{\curvatureVariable}}$ or vice versa; since the relation between the two is a globally invertible function both functional relations are equivalent.

Concretely, we can express $\expval{\op{X}}$ as a function of $\expval{\op{\curvatureVariable}}$,
\begin{equation}
  (
    \expval{\op{X}}
    \circ
    \expval{\op{K}}^{-1}
  )
  (\expval{\op{\curvatureVariable}})
  =
  \expval{\op{X}}(0)
  -
  \frac{2\lambda\expval{\op{\Pi}_1}}{\alpha}
  \log\left(
  \frac{\expval{\op{\curvatureVariable}}}{\expval{\op{\curvatureVariable}}(0)}
  \right)
\end{equation}
 which can be compared with the classical solution we found in \secref{sec:relation_to_classical_friedmann_cosmology},
\begin{equation}
  (
    \varphi
    \circ
    p_V^{-1}
  )
  (p_V)
  =
  \varphi(0)
  \pm
  \frac{1}{\sqrt{12\pi G}}
  \log\left(
    \frac{p_V}{p_V(0)}
  \right)
  \mathperiod
\end{equation}

The classical solution for a flat FLRW cosmology is hence reproduced by the solutions \eqref{eq:flow_expval_scalar_field} and \eqref{eq:flow_expval_curvature_variable} of our effective one-body dynamics if we identify $\expval{\op{X}}$ with a constant multiple of $\varphi$ and $\expval{\op{\curvatureVariable}}$ with a constant multiple of $p_V$. The ratio $\lambda/\alpha$, which is determined by the GFT dynamics and by the operators appearing in the algebra \eqref{algebra}, corresponds to Newton's constant in classical cosmology, with the relation between the two determined by the conserved quantity $\expval{\op{\Pi}_1}$ and the identification between $\expval{\op{X}}$ and the classical $\varphi$. (Recall from the discussion below \eqref{flow2} that we need to define an intensive quantity from the extensive $\expval{\op{X}}$ in order to justify the interpretation as a physical scalar field.) If one identifies the quotient $\expval{\op{X}}/\expval{\op{\Pi}_1}$ of two extensive quantities with the classical scalar field $\varphi$ and the parameters $\lambda$ and/or $\alpha$ are chosen appropriately, arbitrary initial conditions of our effective equations lead to a solution of classical FLRW cosmology. This is the analogue of the recovery of (modified) Friedmann cosmologies in previous work on GFT cosmology such as \cite{Oriti:2016qtz,Oriti:2016ueo,Gielen:2019kae}, where a similar identification of parameters of the fundamental theory with an emergent Newton's constant is made\footnote{That gravitational and, more generally, effective field theory couplings like Newton's constant or the cosmological constant are in fact a function of microscopic parameters of the underlying non-spatiotemporal quantum gravity dynamics is something to be expected in any formalism in which spacetime itself is emergent.}.

An interesting property of the system of equations \eqref{eq:flow_expval_scalar_field}--\eqref{eq:flow_Pi2X_Momentum} is that expectation values and moments are entirely decoupled: the inclusion of  $O(\hbar)$ moments does not alter at all the dynamics of the original variables in $\mathcal{V}$ which correspond to $O(\hbar^0)$. Indeed, the solutions \eqref{eq:flow_expval_scalar_field}--\eqref{eq:flow_expval_curvature_variable} are exactly the ones already discussed in
  \secref{sec:relation_to_classical_friedmann_cosmology} where we fixed the form of the commutator involving $\op{\curvatureVariable}$. This behaviour is due to the linearity of the initial constraint in all variables, and one might see it as representing an unphysical aspect of the truncation we have been using. Indeed, it would seem to imply that semiclassicality properties of the fundamental state are mostly irrelevant to the extraction of interesting cosmological dynamics. We should point out, however, that our results here are fully compatible with previous derivations of effective cosmological Friedmann equations from GFT dynamics as, e.g., given in \cite{Gielen:2019kae}: also in this previous work, effective Friedmann equations can be derived for general states and only depend on expectation values of a few elementary operators, without any dependence on quantum fluctuations or higher moments in general.
  
  The criterion of semiclassicality, as studied in terms of relative fluctuations in \cite{Gielen:2019kae,Marchetti:2020qsq}, should then be seen as an additional requirement for a meaningful physical interpretation of the resulting effective Friedmann equations (derived from the dynamics of expectation values of relevant operators), even if these equations themselves are unaffected by the magnitude of quantum fluctuations. A quantum cosmology with large quantum fluctuations in macroscopic cosmological observables would not describe the universe we observe, even if such observables happen to satisfy the correct (Friedmann-like) dynamics in mean value. In this respect, it is important to emphasise that if relative quantum fluctuations are required to be small at small curvature ($\vert \expval{\op{\curvatureVariable}}(t)/\expval{\op{\curvatureVariable}}(0)\vert \ll 1$), in our context they turn out to be also small when the curvature is large ($\vert \expval{\op{\curvatureVariable}}(t)/\expval{\op{\curvatureVariable}}(0)\vert \gg 1$). Both these limits are characterised by $\vert t\vert\to\infty$, so that the asymptotic behaviour of the relative fluctuations
\begin{equation}
\label{eq:relativefluc}
\sigma^2_{aa'}= \frac{\Delta(aa')}{\expval{\op{a}}\expval{\op{a}'}}
\end{equation}
is identical\footnote{Notice that relative quantum fluctuations are not necessarily small at all times, since they may diverge when the denominator of \eqref{eq:relativefluc} goes to zero. This does not signal that the system is affected by strong quantum fluctuations, but rather that they should not be measured by relative quantities. A more sensible option in this case would be to define a threshold below which fluctuations are considered small, see \cite{Ashtekar:2005dm} for a more detailed discussion.}, as we can see explicitly from \eqref{eq:flow_expval_scalar_field}--\eqref{eq:flow_Pi2X_Momentum}. The fact that quantum fluctuations can be relatively small even when curvature is large is a feature shared with LQC models \cite{Corichi:2011rt,Rovelli:2013zaa}.
A crucial property of our approach is that the (averaged) particle number remains constant, as we have emphasised several times above, in sharp contrast to previous results on quantum fluctuations in GFT cosmology \cite{Gielen:2019kae,Marchetti:2020qsq} where the (average) particle number had a non-trivial relational evolution.\footnote{Notice again that these previous results refer to different notions of relational dynamics in GFT. The number operator in \cite{Gielen:2019kae} is the \virgolette{time-dependent version} of \eqref{eqn:numberoperator}. On the other hand, the (relational) number operator in \cite{Marchetti:2020umh} is the same as in \eqref{eqn:numberoperator}, although relational evolution is defined only for a specific choice of states (and with respect to a different relational clock).} In these works, it was shown that the evolution of the particle number crucially affects the evolution of relative quantum fluctuations, suppressing them at large volume (when $N\gg 1$) and possibly enhancing them at small volume if $N\lesssim 1$. This mechanism of suppression and enhancement of fluctuations is of course absent here, which may be one of the reasons why quantum fluctuations can remain small even in a high curvature regime, provided they are small at low curvature.

\section{Conclusions and discussion}
\label{sec:conclusions}
In this paper we presented a novel way to obtain an effective cosmological dynamics from GFT, motivated by the interpretation of cosmology as quantum gravity hydrodynamics \cite{Gielen:2013naa,Oriti:2016acw}. Contrary to what was done in previous works, this approach does not depend explicitly on a specific choice of states. In order to achieve this, we extended the effective approach developed in \cite{Bojowald:2009jj,Bojowald:2010xp,Bojowald:2010qw} to describe the behaviour of a finite set of collective GFT observables, assumed to capture macroscopic cosmological physics. That is, we extended this approach to a field-theoretic setting. The dynamics of these observables are governed by a one-body constraint, obtained by integrating the microscopic equation of motion of the model. Similarly to \cite{Oriti:2016qtz} and many subsequent works, this model includes matter in the form of a minimally coupled massless scalar field, and is characterised by negligible interactions. This is reflected also at the level of our algebra of one-body operators of interest, which splits into a geometric two-dimensional subalgebra and a four-dimensional one generated by \virgolette{matter observables} and the number operator.

From the algebra of operators and the one-body constraint we obtained a quantum phase space which can be coordinatised by a finite set of expectation values and moments. We considered only moments up to second order in powers of $\hbar$, obtaining an effectively $27$-dimensional space, described by $6$ expectation values and $21$ moments. At this order, the quantum constraint produces $7$ effective constraints, which in turn generate $6$ independent gauge flows in the truncated quantum phase space. We fixed $5$ of these $6$ gauge flows by requiring that the operator $\op{\curvatureVariable}$ that we identify classically with the extrinsic curvature has vanishing moments. The quantum phase space variables related to the other operator generating the geometric subalgebra were then expressed in terms of the remaining variables through the effective constraints. In this way we obtained an effectively reduced phase space parametrised by $15$ variables: $14$ expectation values and moments related to the four-dimensional \virgolette{matter} subalgebra and one expectation value $\expval{\op{\curvatureVariable}}$. We could use the remaining gauge flow to express the evolution of the $14$ non-geometric variables in terms of $\expval{\op{\curvatureVariable}}$. Since $\op{\curvatureVariable}$ by construction has vanishing quantum fluctuations (they are gauge-fixed to zero), this amounts to an effective relational description of the evolution of matter degrees of freedom with respect to the extrinsic curvature clock. This is a refreshing change of perspective for cosmological applications of relational dynamics in quantum gravity, where it is usually matter degrees of freedom that are used as physical clocks (and rods) \cite{Giesel:2012rb}. In the framework we developed, choosing a geometric clock was the most natural choice, since an analogous elimination of matter variables resulting in a reduced quantum phase space of only geometric quantities (and possibly the expectation value of a matter clock) could not be achieved by the methods explained above.

The way relational evolution is obtained in this framework emphasises the idea that in order to define a notion of relational dynamics in an emergent spacetime context as the one we considered here, some prior coarse-graining process is necessary \cite{Marchetti:2020umh}. Here, this coarse-graining is represented by the assumption that the physics of the system is adequately captured by a finite set of macroscopic observables and one collective quantum constraint. This is however still not enough to define a \virgolette{good clock} and consequently a notion of relational dynamics analogous to the classical theory: it is also necessary to be able to find a gauge (the \virgolette{Zeitgeist} of \cite{Bojowald:2010xp,Bojowald:2010qw}) in which quantum fluctuations of the clock can be set to zero.

At the level of expectation values, the resulting relational evolution is in agreement with a classical flat FLRW universe. Interestingly, this result does not depend on the magnitude of quantum fluctuations who evolve independently and, if they are required to be relatively small at low curvature, remain small  at high curvature. We argue that this feature, shared also with LQC models \cite{Corichi:2011rt,Rovelli:2013zaa}, may be due to the constancy of the number operator. In the framework presented in this paper, the number operator is constant because the quantum constraint is a one-body operator and, as such, trivially commutes with the number operator. In most previous attempts to extract cosmological physics from GFT the particle number is time-dependent, and this time dependence can crucially enhance or suppress quantum fluctuations \cite{Marchetti:2020qsq,Gielen:2019kae}.

The methods employed in this work, and thus all the aforementioned results, rely on two important assumptions. 

The first one is related to the choice of states. While we remark again that no specific state choice is necessary within this framework, the states we considered are required to satisfy two important conditions: (i) they should be such that the system is characterised by a few macroscopic observables (hydrodynamic condition); and (ii) moments of operators computed in these states should naturally produce a hierarchy characterised by different powers of $\hbar$ (semiclassicality condition). Without the first requirement one would have to take into account the fact that a field theory is characterised by infinitely many operators and \virgolette{constraints} (the quantum equations of motion), while without the second one one would have to consider the infinitely many moments coordinatising the quantum phase space, even for finite-dimensional systems. 

The second assumption is related to the choice of the fundamental observables, which depends on the features of the physical system being considered. In our case, we introduced the \virgolette{curvature operator} $\op{\curvatureVariable}$ by requiring it to satisfy certain commutation relations which classically allow for a cosmological interpretation of the system. However, a microscopic description of such an operator is missing and requires further studies. In fact, we showed that such an operator cannot be represented on a compact group, suggesting that, at least for the type of dynamics we defined here (which is very much simplified compared to the one corresponding to more realistic quantum gravity models), GFT models with non-compact ``gravitational gauge group'' would be preferred. 

Even when these assumptions are made, there are some general limitations to the framework used here. Computations become more and more involved as either the number of fundamental elements of the quantum algebra increase or the commutation relations between the operators become more complicated (e.g., non-linear). These computational problems are the main reason why considering a system of operators and dynamics for which the particle number can vary is complicated. Indeed, in order to do so, one would have to consider an additional $n$-body constraint ($n>1$) with non-trivial commutation relations with the number operator. An example would be, of course, a dynamical constraint incorporating the contribution from GFT interactions. This would however force us to include many other operators (with possibly more complicated commutation relations) in order to obtain a closed algebra, which would significantly enlarge the quantum phase space.

Despite these limitations, the method we employed allowed us to (i) obtain  effective cosmological dynamics from GFT, (ii) gain insights about the notion of relational dynamics in an emergent quantum gravity scenario, and (iii) study the impact of quantum fluctuations, by making only broad assumptions on the states of interest. The relative simplicity in which the above results were obtained within a well-defined framework illustrates the importance of effective methods for extracting continuum physics from quantum gravity, and in particular suggests that the connection between GFTs (in their hydrodynamic approximation) and cosmological physics may go beyond the choice of specific (condensate) states.

\acknowledgments

The work of SG was funded by the Royal Society through a University Research Fellowship (UF160622) and a Research Grant for Research Fellows (RGF\textbackslash R1\textbackslash 180030).
AP was supported by the same Research Grant for Research Fellows awarded to SG. LM thanks the University of Pisa and the INFN (section of Pisa) for financial support, and the Ludwig Maximilians-Universit\"at (LMU) Munich for the hospitality.

\appendix
\section{Quantum Poisson structure}
\label{app:quantum_poisson_structure}
In this section we collect the non-vanishing Poisson brackets of the quantum phase space considered in \secref{sec:one-body_relational_cosmology}.
In the formulas below we identify $N = \Pi_0$ and introduce the ``fake'' variable $\Pi_{-1} = 0$.
Unless specified otherwise, $a \in \{{N}, {\Pi}_1, {\Pi}_2, {X}, {\areaVariable}, {\curvatureVariable}\}$.

Using \eqref{eq:poisson_brackets_truncated_quantum_phasespace} one can verify the following commutation relations which we computed using Wolfram Mathematica,
\allowdisplaybreaks
\begin{align}
  \poissonbracket{
    \expval{\op{\areaVariable}}
  }{
    \moment{\curvatureVariable a}
  }
  &
  =
  (1 + \delta_{a \curvatureVariable})
  \alpha
  \moment{\curvatureVariable a}
  \mathcomma
  \\
  \poissonbracket{
    \expval{\op{\curvatureVariable}}
  }{
    \moment{\areaVariable a}
  }
  &
  =
  -
  (1 + \delta_{a \areaVariable})
  \alpha
  \moment{\curvatureVariable a}
  \mathcomma
  \\
  \poissonbracket{
    \expval{\op{X}}
  }{
    \moment{\Pi_n a}
  }
  &
  =
  (1 + \delta_{a \Pi_n})
  n
  \moment{\Pi_{n-1} a}
  \mathcomma\quad
  a \neq \Pi_{m \neq n}
  \mathcomma
  \\
  \poissonbracket{
    \expval{\op{X}}
  }{
    \moment{\Pi_n \Pi_m}
  }
  &
  =
  n
  \moment{\Pi_{n-1} \Pi_{m}}
  +
  m
  \moment{\Pi_{n} \Pi_{m-1}}
  \mathcomma
  \\
  \poissonbracket{
    \expval{\op{\Pi}_n}
  }{
    \moment{X a}
  }
  &
  =
  -
  (1 + \delta_{a X})
  n
  \moment{\Pi_{n-1} a}
  \mathcomma
  \\
  \poissonbracket{
    \moment{\areaVariable X}
  }{
    \moment{\curvatureVariable a}
  }
  &
  =
  (1 + \delta_{a \curvatureVariable})
  \alpha
  \expval{\op{\curvatureVariable}}
  \moment{X a}
  \mathcomma\quad
  a \neq \Pi_n
  \mathcomma
  \\
  \poissonbracket{
    \moment{\areaVariable X}
  }{
    \moment{\Pi_n a}
  }
  &
  =
  (1 + \delta_{a \Pi_n})
  n
  \expval{\op{\Pi}_{n-1}}
  \moment{\areaVariable a}
  \mathcomma\quad
  a \neq \curvatureVariable, \Pi_{m \neq n}
  \mathcomma
  \\
  \poissonbracket{
    \moment{\areaVariable X}
  }{
    \moment{\curvatureVariable \Pi_n}
  }
  &
  =
  n
  \expval{\op{\Pi}_{n-1}}
  \moment{\areaVariable \curvatureVariable}
  +
  \alpha
  \expval{\op{\curvatureVariable}}
  \moment{X \Pi_n}
  \mathcomma
  \\
  \poissonbracket{
    \moment{\areaVariable X}
  }{
    \moment{\Pi_n \Pi_m}
  }
  &
  =
  n
  \expval{\op{\Pi}_{n-1}}
  \moment{\areaVariable \Pi_m}
  +
  m
  \expval{\op{\Pi}_{m-1}}
  \moment{\areaVariable \Pi_n}
  \mathcomma
  \\
  \poissonbracket{
    \moment{\curvatureVariable X}
  }{
    \moment{\areaVariable a}
  }
  &
  =
  -
  (1 + \delta_{a \areaVariable})
  \alpha
  \expval{\op{\curvatureVariable}}
  \moment{X a}
  \mathcomma\quad
  a \neq \Pi_n
  \mathcomma
  \\
  \poissonbracket{
    \moment{\curvatureVariable X}
  }{
    \moment{\Pi_n a}
  }
  &
  =
  (1 + \delta_{a \Pi_n})
  n
  \expval{\op{\Pi}_{n-1}}
  \moment{\curvatureVariable a}
  \mathcomma\quad
  a \neq \areaVariable, \Pi_{m \neq n}
  \mathcomma
  \\
  \poissonbracket{
    \moment{\curvatureVariable X}
  }{
    \moment{\areaVariable \Pi_n}
  }
  &
  =
  n
  \expval{\op{\Pi}_{n-1}}
  \moment{\curvatureVariable \areaVariable}
  -
  \alpha
  \expval{\op{\curvatureVariable}}
  \moment{X \Pi_n}
  \mathcomma
  \\
  \poissonbracket{
    \moment{\curvatureVariable X}
  }{
    \moment{\Pi_n \Pi_m}
  }
  &
  =
  n
  \expval{\op{\Pi}_{n-1}}
  \moment{\curvatureVariable \Pi_m}
  +
  m
  \expval{\op{\Pi}_{m-1}}
  \moment{\curvatureVariable \Pi_n}
  \mathcomma
  \\
  \poissonbracket{
    \moment{\areaVariable \Pi_n}
  }{
    \moment{\curvatureVariable a}
  }
  &
  =
  (1 + \delta_{a \curvatureVariable})
  \alpha
  \expval{\op{\curvatureVariable}}
  \moment{\Pi_n a}
  \mathcomma\quad
  a \neq X
  \mathcomma
  \\
  \poissonbracket{
    \moment{\areaVariable \Pi_n}
  }{
    \moment{X a}
  }
  &
  =
  -
  (1 + \delta_{a X})
  n
  \expval{\op{\Pi}_{n-1}}
  \moment{\areaVariable a}
  \mathcomma\quad
  a \neq \curvatureVariable
  \mathcomma
  \\
  \poissonbracket{
    \moment{\areaVariable \Pi_n}
  }{
    \moment{\curvatureVariable X}
  }
  &
  =
  -
  n
  \expval{\op{\Pi}_{n-1}}
  \moment{\areaVariable a}
  +
  \alpha
  \expval{\op{\curvatureVariable}}
  \moment{\Pi_n a}
  \mathcomma
  \\
  \poissonbracket{
    \moment{\curvatureVariable \Pi_n}
  }{
    \moment{\areaVariable a}
  }
  &
  =
  -
  (1 + \delta_{a \areaVariable})
  \alpha
  \expval{\op{\curvatureVariable}}
  \moment{\Pi_n a}
  \mathcomma\quad
  a \neq X
  \mathcomma
  \\
  \poissonbracket{
    \moment{\curvatureVariable \Pi_n}
  }{
    \moment{X a}
  }
  &
  =
  -
  (1 + \delta_{a X})
  n
  \expval{\op{\Pi}_{n-1}}
  \moment{\curvatureVariable a}
  \mathcomma\quad
  a \neq \areaVariable
  \mathcomma
  \\
  \poissonbracket{
    \moment{\curvatureVariable \Pi_n}
  }{
    \moment{\areaVariable X}
  }
  &
  =
  -
  n
  \expval{\op{\Pi}_{n-1}}
  \moment{\curvatureVariable a}
  -
  \alpha
  \expval{\op{\curvatureVariable}}
  \moment{\Pi_n a}
  \mathcomma
  \\
  \poissonbracket{
    \moment{\curvatureVariable^2}
  }{
    \moment{\areaVariable a}
  }
  &
  =
  -
  (1 + \delta_{a \areaVariable})
  2 \alpha
  \expval{\op{\curvatureVariable}}
  \moment{\curvatureVariable a}
  \mathcomma
  \\
  \poissonbracket{
    \moment{\areaVariable^2}
  }{
    \moment{\curvatureVariable a}
  }
  &
  =
  (1 + \delta_{a \curvatureVariable})
  2 \alpha
  \expval{\op{\curvatureVariable}}
  \moment{\areaVariable a}
  \mathcomma
  \\
  \poissonbracket{
    \moment{\areaVariable \curvatureVariable}
  }{
    \moment{\curvatureVariable a}
  }
  &
  =
  \alpha
  \expval{\op{\curvatureVariable}}
  \moment{\curvatureVariable a}
  \mathcomma\quad
  a \neq \areaVariable
  \mathcomma
  \\
  \poissonbracket{
    \moment{\areaVariable \curvatureVariable}
  }{
    \moment{\areaVariable a}
  }
  &
  =
  -
  \alpha
  \expval{\op{\curvatureVariable}}
  \moment{\areaVariable a}
  \mathcomma\quad
  a \neq \curvatureVariable
  \mathcomma
  \\
  \poissonbracket{
    \moment{X^2}
  }{
    \moment{\Pi_n a}
  }
  &
  =
  (1 + \delta_{a \Pi_n})
  2 n
  \expval{\op{\Pi}_{n-1}}
  \moment{X a}
  \mathcomma\quad
  a \neq \Pi_{m \neq n}
  \mathcomma
  \\
  \poissonbracket{
    \moment{X^2}
  }{
    \moment{\Pi_n \Pi_m}
  }
  &
  =
  2
  (
    n
    \expval{\op{\Pi}_{n-1}}
    \moment{\Pi_m X}
    +
    m
    \expval{\op{\Pi}_{m-1}}
    \moment{\Pi_n X}
  )
  \mathcomma
  \\
  \poissonbracket{
    \moment{X \Pi_n}
  }{
    \moment{\Pi_m a}
  }
  &
  =
  (1 + \delta_{a \Pi_m})
  m
  \expval{\op{\Pi}_{m-1}}
  \moment{\Pi_n a}
  \mathcomma\quad
  a \neq X, \Pi_{l \neq m}
  \mathcomma
  \\
  \poissonbracket{
    \moment{X \Pi_n}
  }{
    \moment{\Pi_m \Pi_l}
  }
  &
  =
  m
  \expval{\op{\Pi}_{m-1}}
  \moment{\Pi_n \Pi_l}
  +
  l
  \expval{\op{\Pi}_{l-1}}
  \moment{\Pi_n \Pi_m}
  \mathcomma
  \\
  \poissonbracket{
    \moment{X \Pi_n}
  }{
    \moment{X a}
  }
  &
  =
  -
  (1 + \delta_{a X})
  n
  \expval{\op{\Pi}_{n-1}}
  \moment{\Pi_n \Pi_l}
  \mathcomma\quad
  a \neq \Pi_m
  \mathcomma
  \\
  \poissonbracket{
    \moment{\Pi_n \Pi_m}
  }{
    \moment{X a}
  }
  &
  =
  -
  (1 + \delta_{a X})
  (
    n
    \expval{\op{\Pi}_{n-1}}
    \moment{\Pi_m a}
    +
    m
    \expval{\op{\Pi}_{m-1}}
    \moment{\Pi_n a}
  )
  \mathperiod
\end{align}

\bibliographystyle{jhep}
\bibliography{bibliography.bib}

\end{document}